\RequirePackage[hyphens]{url}
\documentclass[10pt,journal,compsoc]{IEEEtran}
\usepackage{amsmath}
\usepackage{amsthm}
\usepackage{url}
\usepackage{lipsum}
\usepackage{enumitem}
\usepackage{xspace}
\usepackage{subfig}
\usepackage{amsmath}
\usepackage{color}
\usepackage{graphicx}
\usepackage{booktabs}
\usepackage{hyperref}
\usepackage{amssymb}

\usepackage{setspace}
\usepackage{threeparttable}
\usepackage{makecell}
\usepackage{bm}

\usepackage{lscape}
\usepackage{graphicx}
\usepackage{float}
\usepackage{subfig}

\usepackage{multirow}
\usepackage{multicol}
\usepackage{enumitem}
\usepackage{colortbl}
\usepackage[linesnumbered,boxed,ruled,vlined,noend]{algorithm2e}

\usepackage[labelfont=bf, textfont=md, font=small,aboveskip=0pt,belowskip=0pt]{caption}

\newcommand{\vect}[1]{\boldsymbol{#1}}
\newtheorem{theorem}{Theorem}
\newtheorem{lemma}{\textbf{Lemma}}
\newtheorem{problem}{Problem}

\newcommand{\beq}[1]{\small \begin{equation}#1\end{equation}
\normalsize
}

\newcommand{\besp}[1]{\begin{split}#1\end{split}}

\newcommand{\hide}[1]{} 

\newcommand{\qiu}[1]{\textit{\color{red}[(Qiu: #1 )]}}  %

\newcommand{\vpara}[1]{\vspace{0.1in}\noindent\textbf{#1 }}
\newcommand{\model}[0]{SketchNE\xspace}
\newcommand{\revise}[1]{{\color{black}{#1}}}

\def\Figref#1{Fig.~\ref{#1}}

\def\Secref#1{Sec.~\ref{#1}}

\def\eqref#1{eq.~\ref{#1}}

\def\Algref#1{Alg.~\ref{#1}}

\def\1{\bm{1}}

\def\mA{{\bm{A}}}
\def\mB{{\bm{B}}}

\def\mD{{\bm{D}}}

\def\mH{{\bm{H}}}

\def\mK{{\bm{K}}}
\def\mL{{\bm{L}}}
\def\mM{{\bm{M}}}

\def\mO{{\bm{O}}}
\def\mP{{\bm{P}}}
\def\mQ{{\bm{Q}}}
\def\mR{{\bm{R}}}

\def\mU{{\bm{U}}}
\def\mV{{\bm{V}}}

\def\mX{{\bm{X}}}

\def\mZ{{\bm{Z}}}

\def\mLambda{{\bm{\Lambda}}}
\def\mSigma{{\bm{\Sigma}}}

\newcommand{\R}{\mathbb{R}}

\DeclareMathOperator{\tln}{trunc\_log}
\DeclareMathOperator{\htln}{trunc\_log^\circ}

\ifCLASSOPTIONcompsoc
  \usepackage[nocompress]{cite}
\else
  \usepackage{cite}
\fi
\ifCLASSINFOpdf
\else
\fi

\PassOptionsToPackage{hyphens}{url}\usepackage{hyperref}
\begin{document}
%
\title{\model: 
		Embedding Billion-Scale Networks Accurately in One Hour  
}
%
%
%
%

\author{Yuyang~Xie, 
        Yuxiao~Dong,
        Jiezhong~Qiu, 
        Wenjian~Yu,
        Xu~Feng,
        Jie~Tang
\thanks{Manuscript received 26 May 2022; revised 7 Dec. 2022; accepted 23 Feb. 2023.
This work was supported by NSFC under Grant 61872206,  in part by the National Key R\&D Program of China under Grant 2018YFB1402600, in part by NSFC for Distinguished Young Scholar under Grant 61825602, in part by NSFC under Grant 61836013 and 62276148, in part by Zhipu.AI. (Corresponding author: W. Yu.)

The authors (except J. Qiu) are with  the Department of Computer Science and
Technology, BNRist, Tsinghua University, Beijing 100084, China (e-mail: xyy18@mails.tsinghua.edu.cn, yuxiaod@tsinghua.edu.cn, yu-wj@tsinghua.edu.cn, fx17@mails.tsinghua.edu.cn, jietang@tsinghua.edu.cn).
J. Qiu (email: jiezhongqiu@zhejianglab.com) is with Zhejiang Lab, and this work was partially done when he was a PhD candidate at Tsinghua University.
}
}

%
%

\markboth{IEEE Transactions on Knowledge and Data Engineering,~Vol.~0, No.~0, 2023}%
{Shell \MakeLowercase{\textit{et al.}}: Bare Demo of IEEEtran.cls for Computer Society Journals}
\IEEEtitleabstractindextext{%
\begin{abstract}
We study large-scale network embedding with the goal of generating high-quality embeddings for networks with more than 1 billion vertices and 100 billion edges.  
Recent attempts LightNE and NetSMF propose to sparsify and factorize the (dense) NetMF matrix for embedding large networks, where NetMF is a theoretically-grounded network embedding method. 
However, there is a trade-off between their embeddings' quality and scalability due to their expensive memory requirements, making embeddings less effective under real-world memory constraints.  
Therefore, we present the \model
model, a scalable, effective, and memory-efficient network embedding solution developed for a single machine with CPU only. 
The main idea of \model is to avoid the explicit construction and factorization of the NetMF matrix either sparsely or densely when producing the embeddings through the proposed sparse-sign randomized single-pass SVD algorithm.  
We conduct extensive experiments on nine datasets of various sizes for vertex classification and link prediction, demonstrating the consistent outperformance of \model over state-of-the-art 
baselines in terms of both effectiveness and efficiency. 
\model costs only \textbf{1.0 hours} to embed the Hyperlink2012 network with \textbf{3.5 billion} vertices and \textbf{225 billion} edges on a CPU-only single machine with embedding superiority (e.g., a \textbf{282\%} relative HITS@10 gain over LightNE). 
\end{abstract}

\begin{IEEEkeywords}
network embedding, network representation learning, randomized matrix factorization, memory-efficient.
\end{IEEEkeywords}}

\maketitle
\IEEEdisplaynontitleabstractindextext

%
\IEEEpeerreviewmaketitle


\IEEEraisesectionheading{\section{Introduction}\label{sec:introduction}}
\IEEEPARstart{R}{epresentation} learning on graphs has recently provided a new paradigm for modeling real-world networks~\cite{surveyHamiltonYL17}. 
Learning structural representations for networks, i.e., network embedding, aims to map network entities into a latent space. 
The learned entity embeddings have been used to power various billion-scale online services, such as 
DeepWalk~\cite{deepwalk} in Alibaba~\cite{recomalibaba}, 
LINE~\cite{line} in LinkedIn~\cite{linkedin}, metapath2vec~\cite{dong2017metapath2vec} and NetSMF~\cite{netsmf} in Microsoft Academic~\cite{m2v4MAS}, 
PinSage in Pinterest~\cite{pinsage}. 

\begin{figure*}[h]
	\centering
	\includegraphics[width=0.9\linewidth]{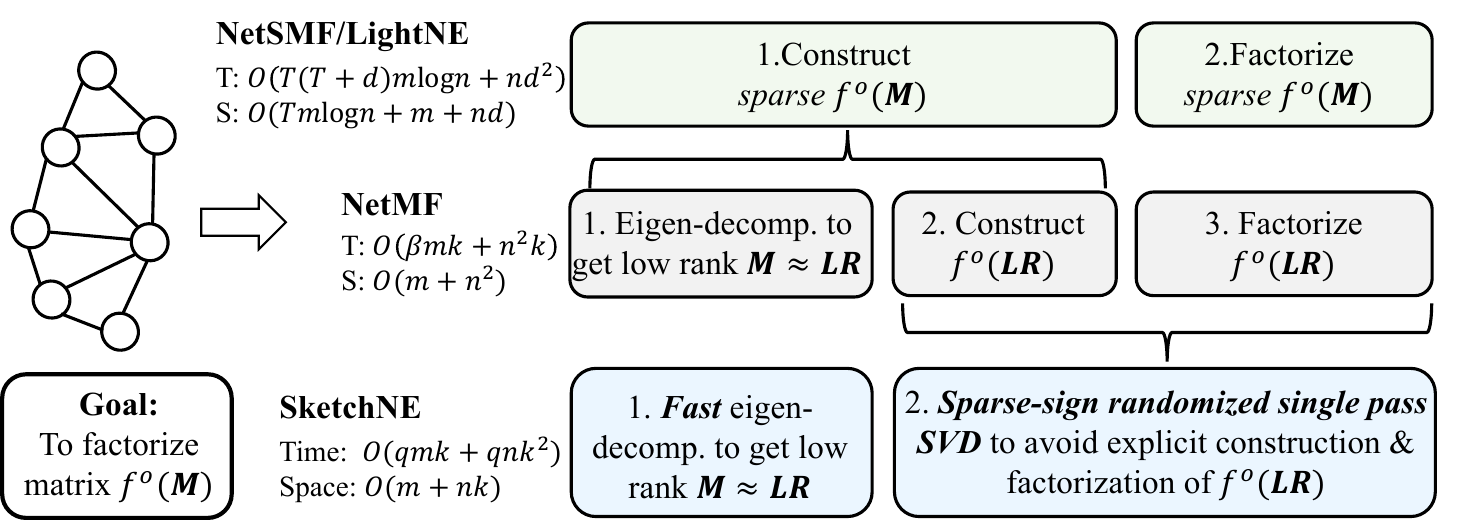}
	\caption{The overview of SketchNE vs. NetMF and NetSMF/LightNE. The symbols used are listed in Table~\ref{tab:notation}.
	}
	\label{fig:simple_overview}
\end{figure*}

Take Facebook for example, it leverages the word2vec~\cite{word2vec} based graph embedding system~\cite{lerer2019pbg} to learn structural embeddings for its 3 billion user base. 
These embeddings are then consumed in various downstream applications. 
To maintain the quality of these embeddings, it is required to periodically embed such networks as its underlying structure consistently evolves, ideally as frequently as possible, 
e.g., every few hours in Alibaba~\cite{recomalibaba}. 
However, according to our estimates, state-of-the-art (SOTA) graph embedding systems, i.e., GraphVite~\cite{graphvite}---a DeepWalk~\cite{deepwalk} based system---and PyTorch-BigGraph~\cite{lerer2019pbg}---would cost days if not weeks by using powerful CPU and GPU clusters to embed a network of 3B users.

Though skip-gram based embedding models, e.g., DeepWalk~\cite{deepwalk}, LINE~\cite{line}, and {metapath2vec~\cite{dong2017metapath2vec}}, have been widely adopted in large-scale solutions. 
They are still limited to handle billion-scale networks at speed, as discussed above. 
Recently, a theoretical study demonstrates that these models can be transformed as implicit factorization of a closed-form matrix~\cite{netmf}.
Based on this discovery, the NetMF model is proposed to explicitly construct and factorize the matrix that is implicitly factorized by DeepWalk, i.e., the NetMF matrix\footnote{The detailed NetMF matrix $f^\circ(\mM)$ can be found in Table~\ref{tab:notation}.} $f^{\circ}(\vect{M})$, in which $\vect{M}$ can be approximated by 
$\vect{L}$ and $\vect{R}$---the two matrices formed by the eigen-decomposition over the graph Laplacian of a given network---and $f^{\circ}(\cdot)$ is an element-wise logarithm function.  
Additionally, addressing 
the matrix form $f^{\circ}(\vect{LR})$ can benefit various machine learning scenarios, such as the attention mechanism in Transformer~\cite{vaswani2017attention}---$\mathrm{softmax}(\cdot)$,
the linear layer with ReLU~\cite{glorot2011relu} activation---$\mathrm{ReLU}(\cdot)$, and the kernel method~\cite{hofmann2008kernel}. 


Despite its  outperformance over skip-gram based methods, it is computationally prohibitive for NetMF to handle million-scale networks as it needs to construct and factorize $f^{\circ}(\vect{M})$, which is an $n\times n$ dense matrix with $n$ being the number of vertices. 
To address this, one recent attempt NetSMF~\cite{netsmf} proposes to construct a sparse version of $f^{\circ}(\vect{M})$ by a graph spectral based sampling technique and then leverage sparse matrix factorization to produce vertex embeddings. 
More recently, LightNE~\cite{lightne} advances NetSMF by further reducing its sampling cost, utilizing other system-wise optimizations, and borrowing the spectral propagation strategy from ProNE~\cite{prone}. 
In doing so, LightNE outperforms SOTA systems, including NetSMF, ProNE, GraphVite, and PyTorch-BigGraph, in terms of both computational cost and embedding effectiveness.

However, the performance of LightNE and NetSMF heavily relies on the number of samplings that directly corresponds to the memory cost, that is, more samplings make the sparse matrix more close to $f^{\circ}(\vect{M})$ and thus yield better embeddings, while consuming more memory. 
For example, to generate competitive embeddings for the OAG data~\cite{lightne} of 67M vertices and 895M edges, LightNE requires 1493GB memory space to have a sufficient number of samples. 
In order to embed larger networks, such as those of billions of vertices, LightNE has to sacrifice the quality of the embeddings under the real-world memory constraint.

\vpara{Contributions.}
In light of the limitations of existing large-scale graph embedding solutions, the goal of this work is to learn effective embeddings for billion-scale networks efficiently under certain memory constraints, e.g., to embed networks with 3B vertices and 200B edges within 1 hour by using a single machine with 1500GB memory.
To achieve this, we present the \model\footnote{The code is publicly available at \url{https://github.com/xyyphant0m/SketchNE}} model, an effective, scalable, and memory-efficient 
method for billion-scale network embedding. 
Figure \ref{fig:simple_overview} illustrates the two technical components in \model---a fast eigen-decomposition algorithm and a sparse-sign randomized single-pass SVD, each of which addresses the computational challenges in NetMF correspondingly.

First, we propose to factorize the target matrix $f^{\circ}(\vect{M})$ without explicitly constructing it, avoiding the direct or sparse construction and factorization in NetMF or NetSMF/LightNE.  
To achieve this, we present the sparse-sign randomized single-pass SVD algorithm by leveraging the concept of the randomized sketch matrix. 

Second, the step above still requires $\vect{L}$ and $\vect{R}$, though the explicit construction of $f^{\circ}(\vect{M})$ is not demanded anymore. 
Thus, we further introduce a fast randomized eigen-decomposition algorithm to approximate the computation of $\vect{L}$ and $\vect{R}$ and give an upper bound of the approximation error. Empirical tests show that we can achieve about 90$\times$ speedup over the original eigen-decomposition module in NetMF without performance loss on (small) networks that NetMF can actually handle. 

Third, we combine the spectral propagation strategy which is proposed in~\cite{prone} to further improve the quality of the inital embedding. We optimize our system for shared-memory architectures with Graph Based Benchmark Suite (GBBS)~\cite{gbbs}, which has already shown its superiority when handling real-world networks with hundreds of billions of edges on a single machine. Intel Math Kernel Library (MKL) is used in \model for basic linear algebra operations.

We conduct extensive experiments to examine the performance of \model, including its effectiveness, efficiency, and memory cost. Specifically, we test \model and other SOTA models/systems on five datasets for vertex classification and four datasets for link prediction. 
The results show that by using the least running time and memory among SOTA models/systems, \model can consistently outperform nine large-scale baselines across five datasets for vertex classification and also offers significant improvements over LightNE on three billion-scale networks for link prediction. 
Notably, \textit{
\model can embed the Hyperlink2012 network with \textbf{3.5 billion vertices} and \textbf{225 billion edges} in \textbf{1.0 hours} by using 1,321GB memory on a single machine, and the learned embeddings offer a \textbf{282\%} relative HITS@10 improvement over LightNE on the link prediction task. 
}

\begin{table}[b]
	\centering
	\caption{Symbol used throughout this paper.}
	\label{tab:notation} 
	\footnotesize
	\begin{tabular}[htbp]{@{}l@{}|@{}r@{~}|@{~}l@{~}|@{}r@{}}
		\toprule
		\textbf{Symbol} & \textbf{Description} & \textbf{Symbol} & \textbf{Description}                    \\  \midrule
		$G$               & input network        & $b$               & \#negative samples                      \\
		$V$               & vertex set, $|V|=n$  & $T$               & context window size                     \\
		$E$               & edge set, $|E|=m$    & $\vect{E}$    &  embedding matrix            \\
		$\vect{A}$        & adjacency matrix     & $d$               & embedding dimension           \\
		$\mathrm{vol}(G)$ & volume of $G$         & $\vect{U}_k$         & truncated eigenvectors            \\
		$\vect{D}$        & degree matrix        & $\vect{\Lambda}_k$ & truncated eigenvalues \\
		$\vect{D}^{\!-\!\alpha}\vect{A}\vect{D}^{\!-\!\alpha}$  & modified Laplacian& $k$ & rank parameter     \\
		$\mathcal{L}$ & normalized  Laplacian& $q$ & power parameter\\
		$\vect{M}$ & $\frac{\mathrm{vol}(G)}{bT}\!\sum_{r=1}^{T}(\vect{D}^{\!-1}\vect{A})^{r}\vect{D}^{\!-1}$ & $\vect{Y}$ & sketch of $f^\circ (\mL \mR)$ \\
		$s$ & oversampling parameter & $z$ & column density\\
		\bottomrule
	\end{tabular}
\end{table}

\section{N\texorpdfstring{\MakeLowercase{et}}{et}MF and Its Challenges}


Given an undirected network $G = (V,E,\vect{A})$ with $n$ vertices, $m$ edges, adjacency matrix $\vect{A}$, degree matrix $\mD$ and volume $\mathrm{vol}(G)=\sum_i\sum_j \mA_{ij}$, the goal of network embedding is to learn an embedding matrix $\vect{E}\in\mathbb{R}^{n\times d}$ so that row $i$ captures the structural property of vertex $i$~\cite{deepwalk,node2vec,line,netmf}. 
The embeddings can be then fed into downstream applications. 
The symbols are listed in Table~\ref{tab:notation}. 


Many network embedding algorithms are based on 
random walk and skip-gram techniques, such as DeepWalk~\cite{deepwalk}, LINE~\cite{line}, and node2vec~\cite{node2vec}. 
Take DeepWalk for example, the vertex sequences traversed by random walkers are fed into the skip-gram model, which is usually parameterized by the context window size $T$ and the number of negative samples $b$. 
Notably, these techniques are later shown to be theoretically equivalent to matrix factorization~\cite{netmf}. 
Based on this result, the NetMF algorithm is proposed to explicitly construct and factorize the matrix that is implicitly factorized by DeepWalk, namely the NetMF matrix:
\beq{
	\label{eq1}
	\mathrm{trunc\_log}^{\circ}\left(\frac{\mathrm{vol}(G)}{bT}\!\sum_{r=1}^{T}(\vect{D}^{-1}\vect{A})^{r}\vect{D}^{\!-1}\right),
}where 
$\mathrm{trunc\_log}^{\circ}$ denotes the element-wise truncated logarithm, i.e., applying $\tln(x)\!=\!\mathrm{max}({0,\mathrm{log}(x)})$ to each entry of a matrix. 
%
However, the explicit construction and factorization of this matrix usually consumes $O(n^3)$ time 
as it tends to be a dense matrix even with a small $T$. 
To reduce time complexity, NetMF conducts truncated eigen-decomposition such that $\vect{D}^{-1/2}\vect{A}\vect{D}^{-1/2} \approx \mU_k \mLambda_k \mU_k^\top$, 
and factorizes the following approximate matrix of (\ref{eq1}):
\beq{
	\label{eq2}
\besp{
		\htln\left(\frac{\mathrm{vol}(G)}{bT}\vect{D}^{-1/2}\vect{U}_k\left(\sum_{r=1}^{T}\vect{\Lambda}_k^{r}\right)\vect{U}_k^{\top}\vect{D}^{-1/2}\right).
}}

\vpara{Reformulate the Goal of NetMF.}
With the above description, we can reformulate and generalize the goal of NetMF as follows:
\begin{problem} Truncated SVD for element-wise function of a low-rank matrix.

\noindent\textbf{Given:} Two low-rank  matrices $\mL \in \R^{n \times k}$ and  $\mR \in \R^{k \times n}$, a function $f: \R \rightarrow \R$ applied to each entry of $\mL \mR$, and desired dimensionality $d$.

\noindent\textbf{Goal:} Compute the rank-$d$ truncated SVD for $f^\circ(\mL \mR)$ such that: 
\beq{
[\mU_d, \mSigma_d, \mV_d] = \mathrm{svds}(f^\circ (\mL \mR), d).
}In NetMF, $\vect{L}\!=\!\frac{\mathrm{vol}(G)}{bT}\vect{D}^{-1/2}\vect{U}_k$, $\vect{R}\!=\!(\sum_{r=1}^{T}\vect{\Lambda}_k^{r})\vect{U}_k^{\top}\vect{D}^{-1/2}$ and $f(\cdot)\!=\!\tln(\cdot)$.
\end{problem}

\hide{
\begin{equation}
	\label{eq3}
	\vect{U},\vect{\Sigma},\vect{V} = \mathrm{svds}(f^{\circ}(\vect{L}\vect{R}),d) ~,
\end{equation} 
where $\vect{L}\!=\!\frac{\mathrm{vol}(G)}{bT}\vect{D}^{-1/2}\vect{U}_k$, $\vect{R}\!=\!(\sum_{r=1}^{T}\vect{\Lambda}_k^{r})\vect{U}_k^{\top}\vect{D}^{-1/2}$,  $f^{\circ}(x)\!=\!\mathrm{trunc\_log}^{\circ}(x)$, and $\mathrm{svds}(,d)$ performs the rank-$d$ truncated SVD. 
}
%
\Algref{alg:netmf_large_window} describes NetMF under the above new formulation. Unfortunately, it  is still not capable of handling large networks, even the YouTube dataset with 1.1 million vertices used in DeepWalk~\cite{deepwalk,youtubedata}, mainly due to the following two challenges presented in Problem 1.


\begin{algorithm}[t!]
    \caption{NetMF under the new formulation}
    \small
    \label{alg:netmf_large_window}
    \SetAlgoLined
    \KwIn{adjacency matrix $\vect{A}$, rank $k$, embedding dimension $d$}
	\KwOut{An embedding matrix $\vect{E}\in\mathbb{R}^{n\times d}$ }
	$[\vect{U}_k,\vect{\Lambda}_k] = \mathrm{eigs}(\vect{D}^{-1/2}\vect{A}\vect{D}^{-1/2},k)$ \tcp*[f]{Eigen-decomposition} \\
    $\vect{L}\!=\!\frac{\mathrm{vol}(G)}{bT}\vect{D}^{-1/2}\vect{U}_k$, $\vect{R}\!=\!\left(\sum_{r=1}^{T}\vect{\Lambda}_k^{r}\right)\vect{U}_k^{\top}\vect{D}^{-1/2}$\\
    $[\vect{U}_d,\vect{\Sigma}_d,\vect{V}_d] = \mathrm{svds}(f^\circ(\mL \mR),d)$ \tcp*[f]{Rank-$d$ truncated SVD} \\
    \Return $\vect{E}=\vect{U}_d \vect{\Sigma}_d^{1/2}$ as network embedding 
\end{algorithm}

\vpara{Challenge One: How to Solve $\mathrm{svds}(f^\circ(\mL \mR), d)$ Efficiently? }
The major challenge lies in the requirement to explicitly construct and factorize  $f^{\circ}(\vect{L}\vect{R})$, even after NetMF's attempt to perform the truncated eigen-decomposition. 
In fact, its construction and factorization in \Algref{alg:netmf_large_window} Line 3 demand $O(n^2)$ memory cost and $O(n^2k)$ time cost, making it computationally infeasible for large networks. 
It is worth noting that the element-wise truncated logarithm is very important to embedding quality and cannot be omitted~\cite{ivashkin2016logarithmic,netmf,netsmf,lightne}. 
Otherwise, the embeddings can be realized without constructing the dense form, as in NRP~\cite{nrp}, RandNE~\cite{randne}, and FastRP~\cite{fastrp}.

\vpara{Challenge Two: How to Factorize $\mD^{-1/2}\mA\mD^{-1/2}$ Efficiently?}
Although one may think the truncated eigen-decomposition of $\mD^{-1/2}\mA\mD^{-1/2}$ (Line 1 of \Algref{alg:netmf_large_window}) is a simple and efficient step,
previous work~\cite{cohen2018approximatingspectrum, liu2013large} and our analysis show that it is in fact  computationally very expensive. 
\hide{
Further, to get $\vect{L}$ and $\vect{R}$ at the first place, the required eigen-decomposition on $\vect{D}^{-1/2}\vect{A}\vect{D}^{-1/2}$, i.e., Line 1 of Alg.~\ref{alg:netmf_large_window}, is computationally very expensive.
}
In particular, Cohen-Steiner et al.~\cite{cohen2018approximatingspectrum} shows it is very difficult to obtain the spectrum of a large graph, and the eigen-decomposition of a large graph Laplacian~\cite{liu2013large} is practically very slow. 
The cost of the truncated eigen-decomposition~\cite{arpack1998} is $O(\beta mk)$, where $\beta\ge1$ and its value depends on the convergence speed. 
The convergence, in turn, depends on the relative eigenvalue gap, 
making this constant term very big and this operation practically very expensive. 
This problem seems not prominent in NetMF because the datasets in its paper~\cite{netmf} are relatively small (the largest one is Flickr with 80K vertices). However, for a slightly larger dataset YouTube with 1.1M vertices, we observe that
the SciPy implementation \texttt{eigsh} cannot finish the computation in three days,
not to mention for billion-scale networks.


\section{The S\texorpdfstring{\MakeLowercase{ketch}}{ketch}NE Model}


In this section, we present \model 
for embedding billion-scale networks at speed. 
In \Secref{subsec:fLR}, we propose a spare-sign single-pass SVD algorithm to resolve challenge one---factorizing $f^{\circ}(\vect{LR})$ without constructing its dense form. 
In \Secref{subsec:fasteigen}, we introduce a fast randomized eigen-decomposition algorithm to resolve challenge two---accelerating the computation of $\mL$ and $\mR$. 
\Secref{subsec:sketchne_overall} combines the two sketching techniques and presents the overall algorithm.


\hide{
\begin{figure*}[ht]
	\setlength{\abovecaptionskip}{0.08 cm}
	\centering
	\includegraphics[width=1\linewidth]{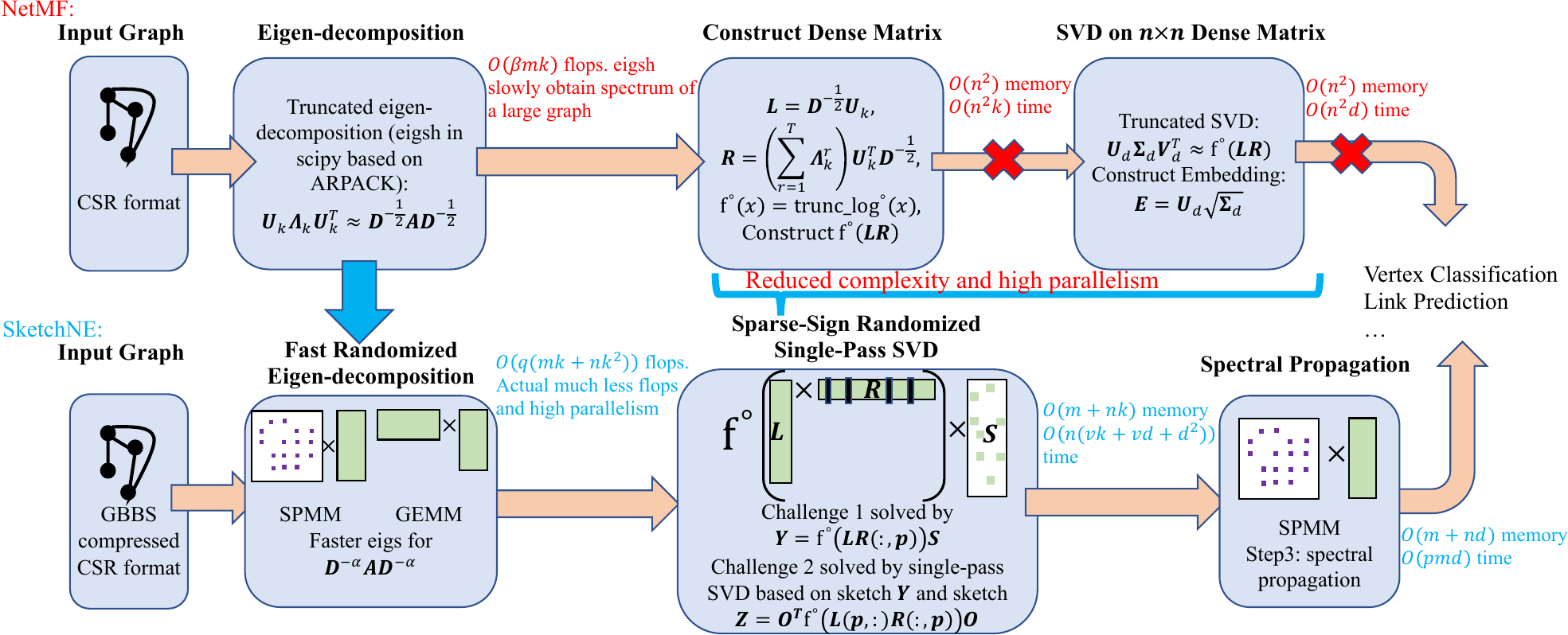}
	\caption{The overview of SketchNE vs. NetMF.}
	\label{fig:sketchnevsnetmf}
\end{figure*}
}


\subsection{Sketching-Based $\mathrm{svds}(f^{\circ}(\mathbf{\emph{LR}}),d)$ without Explicitly Computing and Factorizing $f^\circ ( \mathbf{\emph{LR}})$}
\label{subsec:fLR}


In this part, we formally introduce a sketch-based solution to
$\mathrm{svds}(f^{\circ}(\vect{LR}),d)$ without explicitly computing and factorizing it.

\begin{algorithm}[t]
	\caption{The basic randomized SVD}
	\small
	\label{alg:basic_rsvd}
	\SetAlgoLined\DontPrintSemicolon
	\SetKwFunction{proc}{$\mathrm{basic\_randomized\_SVD}$}
	\SetKwProg{myproc}{Procedure}{}{}
	\myproc{\proc{$\vect{X},k,q$}}{
		$\vect{\Omega}=\mathrm{randn}(n,~k\!+\!s)$\tcp*[f]{oversampling parameter $s$:\>10\>or\>20} \\
    	$\vect{Y}=\vect{X}\vect{\Omega}$ \tcp*[f]{sketch matrix $\R^{n\times (k+s)}$}\\
    	$\vect{Q} = \mathrm{orth}(\vect{Y})$ \tcp*[f]{QR factorization}\\
    	\For(\tcp*[f]{power iteration~(optional) \hide{parameter $q$ }}){$i=1,...,q$}{
    		$\vect{T} = \mathrm{orth}(\vect{X}^{\top}\vect{Q})$ \tcp*[f]{QR factorization}\\
    		$\vect{Q} = \mathrm{orth}(\vect{X}\vect{T})$ \tcp*[f]{QR factorization}\\
    	}
    	$\vect{B}=\vect{Q}^{\top}\vect{X}$ \tcp*[f]{reduced matrix $B\in \R^{(k+s) \times n}$}\\
    	$[\hat{\vect{U}},\hat{\vect{\Sigma}},\hat{\vect{V}}] = \mathrm{svd}(\vect{B})$ \tcp*[f]{Full SVD}\\
    	$\vect{U}=\vect{Q}\hat{\vect{U}}(:,1:k),\vect{\Sigma}=\hat{\vect{\Sigma}}(1:k,1:k),\vect{V}=\hat{\vect{V}}(:,1:k)$\;
		\KwRet $\vect{U},\vect{\Sigma},\vect{V}$\;
	}
\end{algorithm}
\vpara{Basic Randomized SVD.}
To solve the computational challenge of $\mathrm{svds}(f^{\circ}(\vect{LR}),d)$, we first revisit the randomized SVD in \Algref{alg:basic_rsvd}, where $\vect{\Omega}$ (Line 2) is a Gaussian random matrix, 
$s$ (Line 2) is the oversampling parameter for better accuracy, $q$ (Line 5) is the power iteration index, 
and $\mathrm{orth}(\cdot)$ (Lines 4, 6 and 7) is the orthonormalization operation 
which is usually implemented with QR factorization. 
\hide{
These methods use random sampling to identify a subspace that
captures most of the action of a matrix. The input matrix is then compressed—either explicitly or
implicitly—to this subspace, and the reduced matrix is manipulated deterministically to obtain the
desired low-rank factorization.
}
The random projection in Line 3 generates a sketch matrix $\vect{Y}$, which identifies a subspace that captures  dominant information of  the input matrix $\mX\!=\!f^{\circ}(\vect{LR})$.
Then, with the subspace's orthonormal basis matrix $\vect{Q}$ computed in Line 4, one obtains the approximation $\vect{X}\approx\vect{Q}\vect{Q}^{\top}\vect{X}$ ~\cite{halko2011finding}.
Lines 5\raisebox{-0.5ex}{\~{}}8 are the power iteration scheme which is optional to improve the approximation accuracy.
Next, Line 9 constructs a reduced matrix $\vect{B}$
by projecting the input matrix $\mX$ to the subspace with orthonormal basis $\mQ$.
Finally, the approximate truncated SVD of $\vect{X}$ is obtained through performing SVD on matrix $\vect{B}$ in Lines 10-11. 

We can see that there are four places in \Algref{alg:basic_rsvd} (Lines 3, 6, 7 and 9) requiring the explicit construction of $\mX\!=\!f^\circ(\mL \mR)$. 
For Lines 6-7, we can avoid them by  skipping the optional power iteration. 
We   discuss how to deal with issues raised by Lines 3 and 9, respectively.

\hide{
Through this process, the input matrix $\vect{X}$, i.e., $f^{\circ}(\vect{LR})$ is required in both Line 3 and Line 9 for which $\vect{X}$ is also involved in the power iteration in Lines 5\raisebox{-0.5ex}{\~{}}8. 
To avoid its explicit construction, we present solutions to address the computational issues on these two steps. 
}


\vpara{Issue One (\Algref{alg:basic_rsvd} Line 3): Sketching $f^\circ(\mL \mR)$.}
Computing sketch matrix $\vect{Y}$ requires matrix multiplication between $\mX\!=\! f^\circ(\mL \mR)$ and a random matrix  $\vect{\Omega}$. 
However, $f^{\circ}(\vect{LR})$ cannot be explicitly computed due to its $O(n^2k)$ time complexity and $O(n^2)$ memory cost.

\vpara{Solution: The Sparse-Sign Matrix for Sketching.}
We introduce the concept of the sparse-sign matrix to help solve the multiplication $\mX\vect{\Omega}\!=\! f^{\circ}(\vect{LR})\vect{\Omega}$. 
The sparse-sign matrix, with similar performance to the Gaussian matrix, is another type of randomized dimension reduction map~\cite{tropp2019streaming}. 
\Algref{alg:ssmat} describes how to generate a sparse-sign matrix. 
To have a sparse-sign matrix $\vect{S}\in \mathbb{R}^{n\times h}$ with sketch size $h=d+s$, where $d$ is the embedding dimension, and $s$ is the oversampling parameter, we fix a column density parameter $z$ in the range $2\leq z\leq n$. 
We independently generate the columns of the matrix at random. 
For each column, we draw $z$ i.i.d random signs and place them in $z$ uniformly random coordinates as shown in \Algref{alg:ssmat} Lines 5\raisebox{-0.5ex}{\~{}}6. 
Line 8 is to get the unique row coordinates. 
According to~\cite{tropp2019streaming}, $z=\mathrm{min}(n,8)$ will usually be a good choice. 


After generating the sparse-sign matrix $\vect{S}$, we can use it as the random matrix, which multiplies $f^{\circ}(\vect{LR})$ on its right side to obtain the sketch matrix $\vect{Y}$. 
Considering that a column of the sparse-sign matrix $\vect{S}$ only has $z$ nonzeros, it will generate at most $z\times h$ coordinates in the range of $[1,n]$ and $z \times h \ll n$. 
Therefore, we can perform column sampling according to these unique coordinates $\vect{p}$. 
Assuming that $v=\mathrm{size}(\vect{p},1)$, then $v$ satisfy $v\leq z\times h\ll n$. 
Based on the fact that $\vect{S}$ has only $v$ non-zero rows, we can immediately observe that we can have a sampling matrix $\vect{R}(:,~\vect{p})$. 
Therefore, computing $\vect{Y} = f^{\circ}(\vect{LR})\vect{S}$ is exactly equivalent to
\beq{\label{eq4}
	\vect{Y} = f^{\circ}\left(\vect{L}\vect{R}(:,~\vect{p})\right)\vect{S}.
	\vspace{-0.01in}
}The time cost of (\ref{eq4}) is $O(nvk+nvh)$ and the memory cost is $O(nk+nv)$. 
However, when calculating $\vect{L}\vect{R}(:,~\vect{p})$ for a network with billions of vertices, it will introduce $O(nv)$ memory cost, which is still infeasible. To solve this, we adopt the \revise{batch matrix multiplication} by selecting the fixed-size rows of $\vect{L}$ in turn to complete the multiplication, which further reduces the memory cost to $O(nk)$. 


\begin{algorithm}[t!]
	\caption{Generate a sparse-sign matrix}
	\label{alg:ssmat}
	\small
	\SetAlgoLined\DontPrintSemicolon
	\SetKwFunction{proc}{$\mathrm{gen\_sparse\_sign\_matrix}$}
	\SetKwProg{myproc}{Procedure}{}{}
	\myproc{\proc{$n,h,z$}}{
		$\vect{S}=\mathrm{zeros}(n,~h)$\tcp*[f]{sketch size $h$}\\
		$\vect{p} = \mathrm{zeros}(zh,~1)$\tcp*[f]{column density parameter $z$}\\
		\For{$j=1,2,...,h$}{
			$\vect{p}((z(j-1)+1):zj)= \mathrm{randperm}(n,z)$\;
			$\vect{S}(\vect{p}((z(j-1)+1):zj),j) = \mathrm{sign}(\mathrm{randn}(z,1))$\;
		}
		$\vect{p} = \mathrm{unique}(\vect{p})$\tcp*[f]{make coordinates unique}\\
		\KwRet $\vect{S}, \vect{p}$\;
	}
\end{algorithm}

\vpara{Issue Two (\Algref{alg:basic_rsvd} Line 9): Form the Reduced Matrix $\mB$.}
According to \Algref{alg:basic_rsvd} Line 9, the reduced matrix $\vect{B}$ is constructed by  $\vect{B}=\vect{Q}^{\top}f^{\circ}(\vect{LR})$, where $\vect{Q}$ is a dense matrix and $f^{\circ}(\vect{LR})$ is implicitly stored, making it too expensive to obtain $\vect{B}$.

\vpara{Solution: The Randomized Single-Pass SVD.} 
We leverage the idea of randomized single-pass SVD~\cite{tropp2019streaming} to solve this issue. 
The basic idea is to obtain the approximate SVD results by visiting the target matrix $f^{\circ}(\vect{LR})$ only once. 
The process of single-pass SVD is as follows: 
First, we draw multiple sketch matrices that capture the row and column dominant information of matrix $f^{\circ}(\vect{LR})$ and compute SVD based on these sketch matrices. 
In~\cite{tropp2019streaming}, four sparse-sign random matrices 
$\vect{C}\in\mathbb{R}^{n\times h}, \vect{S}\in\mathbb{R}^{n\times h},\vect{H}\in\mathbb{R}^{n\times l}, \vect{O}\in\mathbb{R}^{n\times l}$ are drawn for target matrix $f^{\circ}(\vect{LR})$. 
Then three sketch matrices
\beq{
	\label{eq5}
	\vect{K}=f^{\circ}(\vect{LR})^{\top}\vect{C}, 
	\vect{Y}=f^{\circ}(\vect{LR})\vect{S},
	\vect{Z}=\vect{H}^{\top}f^{\circ}(\vect{LR})\vect{O}
}are generated respectively. 
Second, we obtain the orthonormal matrices
\beq{
    \label{eq6}
	\vect{P}=\mathrm{orth}(\vect{K}),
	\vect{Q}=\mathrm{orth}(\vect{Y}),
}which capture the row and column dominant information of $f^{\circ}(\vect{LR})$, respectively. 
Then we get a great approximation as
\beq{
	\label{eq7}
	f^{\circ}(\vect{LR})\approx\vect{Q}\vect{Q}^{\top}f^{\circ}(\vect{LR})\vect{P}\vect{P}^{\top}.
}By updating $\vect{Z}$ with Eq.~\ref{eq7}, we can have
\beq{
	\label{eq8}
	\mZ =\mH^{\top}f^{\circ}(\mL \mR)\mO \approx (\mH^{\top}\mQ)(\mQ^{\top}f^{\circ}(\mL \mR)\mP)(\mP^{\top}\mO).
}Third, we get the reduced matrix
\beq{
\label{eq9}
\vect{W}=\left(\vect{H}^{\top}\vect{Q}\right)^{\dagger}\vect{Z}\left(\vect{P}^{\top}\vect{O}\right)^{\dagger}
\approx\vect{Q}^{\top}f^{\circ}(\vect{LR})\vect{P}
}by solving the least-squares problem. 
Finally, it will form a low rank approximation of the target matrix $f^{\circ}(\vect{LR})$ via
\beq{
	\label{eq10}
	f^{\circ}(\vect{LR})\approx\vect{Q}\vect{W}\vect{P}^{\top},
}and the approximate truncated SVD of $f^{\circ}(\vect{LR})$ can be derived from performing SVD on $\vect{W}$. Therefore, $f^{\circ}(\vect{LR})$ is required only in the sketching process, and the reduced matrix $\vect{W}$ is constructed only by sketch matrices and random matrices. 

In addition, we note that $f^{\circ}(\vect{LR})$ is symmetric in NetMF, thus the row dominant information is equal to column dominant information which means we can omit $\vect{K}$ and $\vect{H}$. 
This enables us to replace $\vect{P}$ with $\vect{Q}$ in Eq. ~\ref{eq7}\raisebox{-0.5ex}{\~{}}\ref{eq10} and replace $\vect{H}$ with $\vect{O}$ in Eqs. ~\ref{eq5},~\ref{eq8}, and ~\ref{eq9}. 
In other words, we can further simplify and improve the above randomized single-pass SVD process: 
When the multiplication $\vect{Y}=f^{\circ}(\vect{LR})\vect{S}$ is performed, we can simultaneously draw another sketch matrix $\vect{Z}=\vect{O}^{\top}f^{\circ}(\vect{LR})\vect{O}$ with a sparse-sign random matrix $\vect{O}\in\mathbb{R}^{n\times l}$.

\begin{algorithm}[t]
	\caption{\mbox{Sparse-sign randomized single-pass SVD}}
	\label{alg:ssrspSVD}
	\footnotesize
	\SetAlgoLined\DontPrintSemicolon
	\SetKwFunction{proc}{$\mathrm{sparse\_sign\_rand\_single\_pass\_SVD}$}
	\SetKwProg{myproc}{Procedure}{}{}
	\myproc{\proc{$\vect{L},\vect{R},d$}}{
		$h=d+s_1,l=d+s_2$\tcp*[f]{sketch size $h, l$}\\
		$[\vect{S},\vect{p}]=\mathrm{gen\_sparse\_sign\_matrix}(n,h,z)$ \tcp*[f]{\Algref{alg:ssmat}}\\
		$\vect{Y} = f^{\circ}\left(\vect{L}\vect{R}(:,~\vect{p})\right)\vect{S}$\tcp*[f]{sketch matrix $\vect{Y}\in\mathbb{R}^{n\times h}$}\\
		$\vect{Q} = \mathrm{orth}(\vect{Y})$\;
		$[\vect{O},\vect{p}]=\mathrm{gen\_sparse\_sign\_matrix}(n,l,z)$ \tcp*[f]{\Algref{alg:ssmat}}\\
		$\vect{Z} = \vect{O}^{\top} f^{\circ}\left(\vect{L}\left(\vect{p},~:\right)\vect{R}\left(:,~\vect{p}\right)\right)\vect{O}$\tcp*[f]{sketch matrix $\vect{Z}\in\mathbb{R}^{l\times l}$}\\
		$\vect{W}=(\vect{O}^{\top}\vect{Q})^{\dagger}\vect{Z}(\vect{Q}^{\top}\vect{O})^{\dagger}$\tcp*[f]{reduced matrix $\vect{W}\in\mathbb{R}^{h\times h}$}\\
		$[\hat{\vect{U}},\hat{\vect{\Sigma}},\hat{\vect{V}}] = \mathrm{svd}(\vect{W})$ \tcp*[f]{Full SVD}\\ 
		$\vect{U}=\vect{Q}\hat{\vect{U}}(:,1:d),\vect{\Sigma}=\hat{\vect{\Sigma}}(1:d,1:d),\vect{V}=\vect{Q}\hat{\vect{V}}(:,1:d)$\;
		\KwRet $\vect{U},\vect{\Sigma},\vect{V}$\;
	}
\end{algorithm}

\vpara{Overall: Sparse-Sign Randomized Single-Pass SVD.}
By combining the sparse-sign random matrix with single-pass SVD, we propose a sparse-sign randomized single-pass SVD algorithm to avoid the explicit construction and factorization of $f^{\circ}(\vect{LR})$ as \Algref{alg:ssrspSVD}. 

In \Algref{alg:ssrspSVD}, Line 8 generates the reduced matrix $\vect{W}$, which involves solving the least-squares problem twice. 
The first is to solve $(\vect{O}^{\top}\vect{Q})\vect{T}=\vect{Z}$ for the temporary matrix $\vect{T}$ and the second is to solve $(\vect{O}^{\top}\vect{Q})\vect{W}^{\top}=\vect{T}^{\top}$ for the reduced matrix $\vect{W}$. 
The matrix $\vect{O}^{\top}\vect{Q}\in\mathbb{R}^{l\times h}$ is well-conditioned when $l\gg h$ which suggests choosing $s_2 \gg s_1$.
Lines 3 and 6  generate two sparse-sign random matrices. 
Considering that $s_1$ and $s_2$ are small numbers, the time cost of Line 4 is $O(nv(k+d))$ and Line 7 costs $O(v^2(k+d)+vd^2)$. 
Line 5 costs $O(nd^2)$, Lines 8\raisebox{-0.5ex}{\~{}}9 cost $O(d^3)$, and Line 10 costs $O(nd^2)$. 
Therefore, the time complexity of \Algref{alg:ssrspSVD} is $O(n(vk+vd+d^2))$, linear to the number of vertices $n$, which is much more efficient than the explicit construction and factorization in NetMF ($O(n^2k)$). 

Overall, we present the sparse-sign randomized single-pass SVD to address the computational challenges in NetMF, which is the first attempt to introduce  single-pass low-rank matrix factorization into network embedding. 
It not only solves the challenges of NetMF, but also gives a solution to the general problem of factorizing $f^{\circ}(\vect{LR})$.
Recently, Han et al.~\cite{han2020polynomial} proposes polynomial tensor sketch for this problem, which combines a polynomial
approximation of $f$ (e.g., Taylor and Chebyshev expansion) with tensor sketch for approximating monomials of entries
of $\mL\mR$.  
We will see an ablation study which applies polynomial tensor sketch to NetMF in \Secref{exp:ablation}.

\hide{
approximating the element-wise function of low rank matrix based on combining a polynomial approximation of function $f$~(eg. $f(\cdot)=\mathrm{exp}(\cdot)$) with the existing tensor sketch scheme for approximating monomials of entries of $\vect{LR}$.
Therefore, this technique relies on the promise that function $f$ should be a continuous smooth curve for several known (truncated) polynomial expansions such as Taylor or Chebyshev. However, the proposed sparse-sign randomized single-pass SVD is not limited to these and can approximate the non-smooth element-wise function of low rank matrix like $\htln(\vect{LR})$.
}

\subsection{Fast Eigen-Decomposition via Sketching}
\label{subsec:fasteigen}

By now, we bypass the major bottleneck of \Algref{alg:netmf_large_window} (Lines 3) without explicitly computing 
$f^{\circ}(\vect{LR})$. 
However, the solution in \Algref{alg:ssrspSVD} still requires the separate $\vect{{L}}$ and $\vect{{R}}$ as input, which are computed by the eigen-decomposition on $\vect{D}^{-1/2}\vect{A}\vect{D}^{-1/2}$ in \Algref{alg:netmf_large_window} Line 1. 
Though the truncated eigen-decomposition costs only $O(\beta mk)$ FLOPs in theory ($\beta\ge 1$) ~\cite{arpack1998}, it is in practice almost infeasible to handle large networks due to the big constants in its complexity. 
In fact, the computation of this step for the YouTube dataset with 1.1 million vertices cannot complete within three days by using the commonly-used \texttt{eigsh} implementation, while the goal of this work is to embed billion-scale networks efficiently, e.g., in one hour.

To address this practical challenge, we introduce a fast randomized eigen-decomposition method to approximate $\vect{D}^{-1/2}\vect{A}\vect{D}^{-1/2}$. 
According to~\cite{halko2011finding}, the symmetric approximation formula should be $\vect{X}\approx\vect{Q}\vect{Q}^{\top}\vect{X}\vect{Q}\vect{Q}^{\top}$ and the truncated eigen-decomposition result of $\vect{X}$ can be derived by performing eigen-decomposition on the small matrix $\vect{Q}^{\top}\vect{X}\vect{Q}$. 
By combining the techniques of the power iteration scheme and acceleration strategy~\cite{frpca}, the fast randomized eigen-decomposition can be described as \Algref{alg:frevd}.
\begin{algorithm}[t]
	\caption{\mbox{Fast randomized eigen-decomposition}}
	\label{alg:frevd}
	\footnotesize
	\SetAlgoLined\DontPrintSemicolon
	\SetKwFunction{proc}{$\mathrm{freigs}$}
	\SetKwProg{myproc}{Procedure}{}{}
	\myproc{\proc{$\vect{X},k,q$}}{
		$[\sim,n]=\mathrm{size}(\vect{X},~2)$\;
		$\vect{\Omega}=\mathrm{randn}(n,~k+s)$\;
		$\vect{Y}=\vect{X}\vect{\Omega}$\tcp*[f]{sketch matrix $\vect{Y}\in\mathbb{R}^{n\times (k+s)}$}\\
		$[\vect{Q},\sim,\sim] = \mathrm{eigSVD}(\vect{Y})$\tcp*[f]{fast orthonormalization}\\
		\For(\tcp*[f]{power iteration \hide{parameter $q$ }}){$i=1,2,...,q$}{
			$[\vect{Q},\sim,\sim] = \mathrm{eigSVD}(\vect{X}\vect{X}\vect{Q})$\;
		}
		$\vect{S}=\vect{Q}^{\top}\vect{X}\vect{Q}$\tcp*[f]{reduced matrix $\vect{S}\in\mathbb{R}^{(k+s)\times(k+s)}$}\\
		$[\hat{\vect{U}},\vect{\Lambda}] = \mathrm{eig}(\vect{S})$\tcp*[f]{Full eigen-decomposition}\\
		$\vect{U}=\vect{Q}\hat{\vect{U}}$\;
		\KwRet $\vect{U}(:,1:k),\vect{\Lambda}(1:k,1:k)$\;
	}
\end{algorithm}


Practically, a good decomposition of $\vect{D}^{-1/2}\vect{A}\vect{D}^{-1/2}$ by \texttt{freigs} requires a large $q$, increasing the time cost (see the experiment in \Secref{exp:ablation}). 
To balance the trade-off between effectiveness and efficiency, we propose to perform \Algref{alg:frevd} on a modified Laplacian matrix $\vect{D}^{-\alpha}\vect{A}\vect{D}^{-\alpha}$, where $\alpha\in(0,0.5]$. 
Therefore, we have $\vect{U}_k\vect{\Lambda}_k\vect{U}_k^{\top}\approx\vect{D}^{-\alpha}\vect{A}\vect{D}^{-\alpha}$. 
It means $\vect{D}^{-1/2}\vect{A}\vect{D}^{-1/2}$ is computed as $\vect{D}^{-1/2+\alpha}\vect{U}_k\vect{\Lambda}_k\vect{U}_k^{\top}\vect{D}^{-1/2+\alpha}$ approximately. 
We give an upper bound of the approximation error by Lemma 4 and its proof both in Appendix A.
In doing so, Eq. (\ref{eq1}) can be approximated by 
\beq{
	\label{eq11}
	f^{\circ}(\vect{L^{\prime}R^{\prime}}) =  f^{\circ}\left(\frac{\mathrm{vol}(G)}{bT}\vect{D}^{-1+\alpha}\vect{U}_k\vect{\Lambda}_k\left(\sum_{r=1}^{T}\vect{K}^{r-1}\right)\vect{U}_k^{\top}\vect{D}^{-1+\alpha}\right),
}where 
$\vect{L}^{\prime}=\frac{\mathrm{vol}(G)}{bT}\vect{D}^{-1+\alpha}\vect{U}_k$, $\vect{K}=\vect{U}_k^{\top}\vect{D}^{-1+2\alpha}\vect{U}_k\vect{\Lambda}_k$,
$\vect{R}^{\prime}=\vect{\Lambda}_k(\sum_{r=1}^{T}\vect{K}^{r-1})\vect{U}_k^{\top}\vect{D}^{-1+\alpha}$ and
$f(\cdot)=\tln(\cdot)$.
$\vect{K}$ is a $k\times k$ matrix, making the computation of $\sum_{r=1}^{T}\vect{K}^{r-1}$ cheap. 
We further give an upper bound of the approximation error between the NetMF matrix and $f^{\circ}(\vect{L^{\prime}R^{\prime}})$ by the following theorem. 
We can see  the approximation is better with a larger $k$.
\begin{theorem}
Suppose $f^{\circ}$ denotes $\mathrm{trunc\_log}^{\circ}$, i.e. the element-wise truncated logarithm, $f^{\circ}(\vect{M})$ is the matrix in (1), and $f^{\circ}(\vect{L^{\prime}R^{\prime}})$ is defined by (11) which includes the quantities obtained with Alg. 5. Then,\\
    \begin{equation*}
        \Vert f^{\circ}(\vect{M})\!-\!f^{\circ}(\vect{L^{\prime}R^{\prime}})\Vert_F\!\leq\!\frac{(1+\varepsilon)d_{min}^{-1+2\alpha}B}{(c-1)bT}
    \end{equation*}
	with high probability. Here $|\lambda_j|$ is the $j$-th largest absolute value of eigenvalue of $\vect{D}^{-\alpha}\vect{A}\vect{D}^{-\alpha}$, $B=\sqrt{\sum_{j=k+1}^{n}|\lambda_j|^2}((c^T\!-\!1)(1+\frac{n}{c\!-\!1})\!-\!nT)\mathrm{vol}(G)$, $c=n(\frac{d_{max}}{d_{min}})^{1-2\alpha}$. $d_{min}$ and $d_{max}$ are the minimum and maximum vertex degrees, respectively.
\end{theorem}
\begin{proof}
See Appendix A.
\end{proof}


In \Algref{alg:frevd}, the "$\mathrm{eigSVD}$" is used as the orthonormalization operation. 
Compared with the QR factorization, $\mathrm{eigSVD}$ is much faster especially when $n\gg(k+s)$~\cite{frpca}. 
Since the oversampling parameter $s$ is smaller than $k$, Lines 4\raisebox{-0.5ex}{\~{}}5 cost $O(mk+nk^2)$. 
According to~\cite{frpca}, Lines 6\raisebox{-0.5ex}{\~{}}8 cost $O(q(mk+nk^2))$, Line 9 costs $O(mk+nk^2)$, Line 10 costs $O(k^3)$, and Line 11 costs $O(nk^2)$. 
Overall, the time complexity of \Algref{alg:frevd} is $O(q(mk+nk^2))$. 
However, the actual FLOPs of \Algref{alg:frevd} is far fewer than that of \texttt{eigsh}. 
In practice, our empirical tests suggest that by setting $\alpha$=0.4, \Algref{alg:frevd} with a small $q$ shows on average \raisebox{-0.5ex}{\~{}}90X speedup to \texttt{eigsh} on the small datasets that can be handled by \texttt{eigsh} without noticeable impacts on the eigenvalues computed and by extension on the embeddings learned (See \Figref{fig:freigs}~(a)).

\subsection{The Overall Algorithm}
\label{subsec:sketchne_overall}

In \Secref{subsec:fLR}, we develop a sparse-sign randomized single-pass SVD algorithm (\Algref{alg:ssrspSVD}) to solve  $\mathrm{svds}(f^{\circ}(\vect{LR}), d)$ without the explicit computation and factorization of the full matrix $f^{\circ}(\vect{LR})$. 
In \Secref{subsec:fasteigen}, we propose a fast randomized eigen-decomposition method to get $\vect{L}$ and $\vect{R}$ for large networks. 
Empowered by these two techniques, we address the two computational challenges faced by NetMF, respectively. 
To this end, we present the \model algorithm to learn embeddings for billion-scale networks 
in \Algref{alg:SketchNE}.

In \Algref{alg:SketchNE}, Line 1 computes the fast randomized eigen-decomposition for the modified Laplacian matrix $\vect{D}^{-\alpha}\vect{A}\vect{D}^{-\alpha}$. 
Lines 2\raisebox{-0.5ex}{\~{}}3 form the approximations of matrices $\vect{L}$ and $\vect{R}$. 
Line 4 is to compute the SVD of $f^{\circ}(\vect{L^{\prime}R^{\prime}})$ by $\vect{L}^{\prime}$ and $\vect{R}^{\prime}$ through the sparse-sign randomized single-pass SVD. Then we form the inital embedding in Line 5.
Line 6 conducts spectral propagation, which is a commonly-used and computationally-cheap enhancement technique~\cite{prone,lightne} that further improves embedding quality. $c_r$ in Line 6 is the coefficients of Chebyshev polynomials, $p$ is spectral propagation steps (the default setting is 10) and $\mathcal{L}=\vect{I}-\vect{D}^{-1}\vect{A}$ is normalized graph Laplacian matrix ($\vect{I}$ is the identity matrix).
\begin{algorithm}[t]
	\caption{\model}
	\label{alg:SketchNE}
	\footnotesize
	\SetAlgoLined
	\KwIn{A network $G=(V,E,\vect{A})$; Normalized parameter $\alpha$; rank parameter $k$; power parameter $q$; Embedding dimension $d$; }
	\KwOut{An embedding matrix $\vect{E}\in \mathbb{R}^{n\times d}$}
	$[\vect{U}_k,\vect{\Lambda}_k] = \mathrm{freigs}(\vect{D}^{-\alpha}\vect{A}\vect{D}^{-\alpha},k,q)$\tcp*[f]{\Algref{alg:frevd}}\\
	$\vect{K}=\vect{U}_k^{\top}\vect{D}^{-1+2\alpha}\vect{U}_k\vect{\Lambda}_k$\\
	$\vect{L}^\prime=\frac{\mathrm{vol}(G)}{bT}\vect{D}^{-1+\alpha}\vect{U}_k,\vect{R}^\prime=\vect{\Lambda}_k\left(\sum_{r=1}^{T}\vect{K}^{r-1}\right)\vect{U}_k^{\top}\vect{D}^{-1+\alpha}$\\
	$[\vect{U}_d,\vect{\Sigma}_d,\sim]\!=\!\mathrm{sparse\_sign\_rand\_single\_pass\_SVD}(\vect{L}^\prime,\vect{R}^\prime,d)$\tcp*[f]{\Algref{alg:ssrspSVD}}\\
	$\vect{E}=\vect{U}_d\vect{\Sigma}_d^{1/2}$ \tcp*[f]{inital embedding}\\
	$\vect{E}=\sum_{r=0}^{p}c_r\mathcal{L}^{r}\vect{E}$\tcp*[f]{spectral propagation}\\
	\KwRet $\vect{E}$ as network embedding 
\end{algorithm}

\vpara{Complexity of \model.}
For Line 1, the input matrix $\vect{D}^{-\alpha}\vect{A}\vect{D}^{-\alpha}$ is still an $n\times n$ sparse matrix and has $2m$ nonzeros. 
According to Section \ref{subsec:fasteigen}, it requires $O(q(mk+nk^2))$ time and $O(m+nk)$ space. 
As for Lines 2\raisebox{-0.5ex}{\~{}}3, $O(nk^2)$ time and $O(nk)$ space are required. 
The time cost of Line 4 is $O(n(d^2+vk+vd))$ and its space cost is $O(n(k+d))$. 
Lines 5\raisebox{-0.5ex}{\~{}}6 demand $O(pmd+nd)$ time and $O(m+nd)$ space. 
In total, the \model has the time complexity of $O(q(mk+nk^2)+nd^2+nvk+nvd+pmd)$ and the space complexity of $O(m+nk)$. 
Therefore, there is a trade-off between efficiency and effectiveness on the choice of $q$. 
In practice, we can easily find $q$ that offers clear superiority on both efficacy and efficiency, including both memory cost and  computing time, over existing large-scale network embedding techniques. \revise{Consider that $q,k,d,v,p$ are all very small compared to $m$ and $n$, the overall time complexity of \model is linear to the number of edges $m$ and the number of vertices $n$.}

\subsection{Implementation Details}
\vpara{Memory Reduction.}
Considering the memory cost of SketchNE is $O(m+nk)$, while NetSMF/LightNE ties performance to memory cost. Therefore, we consider to further optimize the memory cost of SketchNE by the Graph Based Benchmark Suite (GBBS)~\cite{gbbs}, which is an extension of the Ligra~\cite{ligra} interface. 
We optimize the memory cost of SketchNE with the Graph Based Benchmark Suite (GBBS)~\cite{gbbs}, which is an extension of the Ligra~\cite{ligra} interface. 
The GBBS is easy to use and has already shown its practicality for many large scale fundamental graph problems. LightNE~\cite{lightne} introduces GBBS to network embedding problems and shows its superiority to real-world networks with hundreds of billions of edges.
The main benefit of GBBS to SketchNE is the data compression. 
A sparse adjacency matrix is usually stored in the compressed sparse row (CSR) format, which is also regarded as an excellent compressed graph representation~\cite{kepner2011graph}. However, the CSR format  still incurs a huge memory overhead for the networks with hundreds of billions of edges. For example, storing a network with 1 billion vertices and 100 billion edges  costs 1121 GB  memory. Therefore, we need to compress it further and reduce memory cost. 
The GBBS can be regarded as a compressed CSR format for the graph from Ligra+~\cite{ligraplus}, which supports fast parallel graph encoding and decoding. 

\vpara{Parallelization.}
Two major computational steps of the SketchNE are sparse matrix-matrix multiplication (SPMM) and matrix-matrix product (GEMM), which are well supported by the Intel MKL library. 
SPMM operation is well supported by MKL's Sparse BLAS Routine. 
However, MKL's Sparse BLAS Routine requires the sparse matrix in CSR format as the input, which contradicts the original intention of using GBBS. Fortunately, GBBS supports traversing all neighbors of a vertex $u$ for the compressed CSR format, and we can propose an SPMM operation with the help of GBBS. In order to use GBBS to save memory cost, we propose a parallel GBBS-based SPMM operation to replace the SPMM operation in MKL's sparse BLAS routine.
The parallel GBBS-based SPMM is implemented as follows. Firstly, we traverse $n$ vertices parallelly. Then, we traverse neighbor vertex $v$ of vertex $u$ to compute the quantity $\vect{D}(u,u)^{-\alpha}\vect{D}(v,v)^{-\alpha}$ corresponding to the sparse matrix. Finally, with the support of \texttt{cblas\_saxpy} in MKL, we multiply the $v$-th row of the row-major matrix with the quantity and add the result to the $u$-th row of the target matrix. 
The SPMM operation based on GBBS is slightly slower than MKL's SPMM operation, but ensuring memory-efficient. 
The "eigSVD", "eig" and other operations in SketchNE are well supported by Intel MKL routines. Line 4 and Line 7 of the \Algref{alg:SketchNE} involve matrix column sampling and batch GEMM operation. 
They are  easily parallelized with the "parallel for" derivative in OpenMP~\cite{openmp}.

In conclusion, SketchNE is implemented in C++. We use GBBS to reduce memory usage and implement a GBBS-based SPMM operation. For better efficiency, we use the Intel MKL library for basic linear algebra operations and the OpenMP programming.

\section{Experiments}
\revise{
\begin{table*}[h]
	\caption{\revise{Statistics of datasets.}}
	\label{tab:datasets}
	\centering
	\small
	\renewcommand\arraystretch{0.9}
	{\color{black}\begin{tabular}{@{}r@{}|@{~}r@{~}|@{}r@{~}|@{}r@{~}|@{}r@{~}|@{~}r@{~}|@{}r@{~}|@{}r@{~}|@{}r@{~}|@{}r@{~}}
		\toprule
		&\multicolumn{5}{@{~}c@{~}|}{Multi-label Vertex Classification Task} & \multicolumn{4}{@{~}c@{~}}{Link Prediction Task} \\
		\midrule
		& BlogCatalog & YouTube & Friendster-small & Friendster& OAG& Livejournal & ClueWeb & Hyperlink2014 & Hyperlink2012\\
		\midrule
		$|V|$ & 10,312& 1,138,499& 7,944,949& 65,608,376 & 67,768,244 &4,847,571 &978,408,098 &1,724,573,718 & 3,563,602,789\\
		$|E|$ & 333,983& 2,990,443& 447,219,610& 1,806,067,142 & 895,368,962 &68,993,773 &74,744,358,622 &124,141,874,032 & 225,840,663,232 \\
		\bottomrule
	\end{tabular}}
	\vspace{-0.1in}
\end{table*}
}

In this section, we evaluate \model on multi-label vertex classification and link prediction tasks, 
following exactly the same experimental settings as existing  studies~\cite{deepwalk,node2vec,line,netmf,netsmf,graphvite,lerer2019pbg,lightne}. 
We introduce datasets in Section~\ref{exp:datasets} our experimental settings and results in Section~\ref{exp:setting} and Section~\ref{exp:results}, respectively. The ablation and case studies is in Section~\ref{exp:ablation}.

\subsection{Datasets}
\label{exp:datasets}
We employ five datasets for the multi-label vertex classification task. BlogCatalog and YouTube are small graphs with less than 10 million edges, while the others are large graphs with more than 10 million but less than 10 billion edges.
For the link prediction task, We have four datasets in which vertex labels are not available. Livejournal is the large graph, while the others are very large graphs with more than 10B edges. 
These datasets are of different scales and but have been widely used in network embedding literature~\cite{netsmf,lightne}. 
The statistics of datasets are listed in Table~\ref{tab:datasets}. 

\vpara{BlogCatalog~\cite{blogdata}} is a network of relationships of online
users. The labels represent the interests of the users.

\vpara{YouTube~\cite{youtubedata}} is a video-sharing website, which allows user to
upload, view, rate and share videos. The vertex labels represent the user's taste in the video.

\vpara{Friendster-small~\cite{friendsterdata}} is a sub-graph induced by all the labeled
vertices in Friendster. The vertex labels in this network are the same as those in
Friendster.

\vpara{Friendster~\cite{friendsterdata}} is a large social network in an online
gaming site. For some of the vertices, they have labels representing the groups the user joined.

\vpara{OAG~\cite{oagdata}} is a publicly available academic graph opened by Microsoft Academic~\cite{oagdata} and AMiner.org~\cite{aminer}. The vertex labels represent the study fields of each author. 

\vpara{Livejournal~\cite{livejournaldata}} is an online blogging site, where users can follow others to form a large social network. 

\vpara{ClueWeb~\cite{cluewebdata}} was created to support research on information retrieval and related human language technologies. The links between webs form the very large graphs.

\vpara{Hyperlink2014~\cite{hyperlinkdata}} was extracted from the Common Crawl Corpus released in April 2014, covering 1.7 billion web pages and 124 billion hyperlinks between these pages.

\vpara{Hyperlink2012~\cite{hyperlinkdata}} was extracted from the 2012 Common Crawl Corpus covering 3.5 billion web pages and 225 billion hyperlinks between these pages.

\subsection{Experimental Settings}
\label{exp:setting}
\hide{
We use five datasets  for multi-label vertex classification and four datasets for link prediction.
These datasets are of different scales and  
the largest Hyperlink2012 dataset \cite{hyperlinkdata} is the hyperlinked network from the Common Crawl web corpus, containing 3.5 billion vertices and 225 billion edges. 
The statistics of datasets are listed in Table~\ref{tab:datasets}. 
}

\vpara{Baselines and Hyper-parameters Setting.} 
We compare \model with nine SOTA network embedding methods, including PyTorch-BigGraph (PBG)~\cite{lerer2019pbg},  GraphVite~\cite{graphvite}, NetMF~\cite{netmf}, NetSMF~\cite{netsmf}, ProNE~\cite{prone}, LightNE~\cite{lightne}, RandNE~\cite{randne}, FastRP~\cite{fastrp} and NRP~\cite{nrp}. \revise{We also compare \model with four GNN methods, including DGI~\cite{dgi}, GraphCL~\cite{you2020graphcl}, GCC~\cite{qiu2020gcc} and GraphSAGE~\cite{hamilton2017inductive}.}
For all the baselines originally run on CPU and \model, we test them with all the datasets with 88 threads on a server with two Intel\textsuperscript{\textregistered} Xeon\textsuperscript{\textregistered} E5-2699 v4 CPUs (88 virtual cores in total) and 1.5 TB memory. 
For GraphVite, we present the results obtained from the original paper (if existed), which uses a 4$\times$P100 GPU server, and otherwise run it on a 4$\times$V100 GPU server to get the results. \revise{For GNN methods, we run it on a server with a GeForce GTX 1080 Ti GPU to get the results.} All the baselines are evaluated with the hyperparameters set default in the corresponding paper's GitHub Repository or tuned for the best performance. Their settings are as follows.

\noindent\textbf{NetMF~\cite{netmf}.}
We download the authors' official source codes\footnote{\url{https://github.com/xptree/NetMF}.}, and run experiments with default setting:  $T=10$, $k=256$.

\noindent\textbf{RandNE~\cite{randne}}.
We download the authors' official source codes\footnote{\url{https://github.com/ZW-ZHANG/RandNE}.}, and follow the default hyper-parameter setting for BlogCatalog. 
For other datasets, we follow the suggestion of tuning hyper-parameters from the source codes. The order is from 1 to 3, and the weights are searched according to $w_{i+1}$ = $\beta_iw_i$ where $\beta_i$ is from $\{0.01,0.1,1,10,100\}$.

\noindent\textbf{FastRP~\cite{fastrp}.}
We download the authors' official source codes\footnote{\url{https://github.com/GTmac/FastRP}.}, and follow the authors' suggestion for hyper-parameter setting. $\alpha_1, \alpha_2$ and $\alpha_3$ are set to $0, 0$ and $1$, respectively. We use the official tuning script to tune $\alpha_4$ and the normalization strength $\beta$. The search ranges for $\beta$ and $\alpha_4$ are  $[-1, 0]$ and $[2^{-3}, 2^{6}]$, respectively. 

\noindent\textbf{NRP~\cite{nrp}.}
We download the authors' official source codes\footnote{\url{https://github.com/AnryYang/NRP-code}.}, and follow setting in~\cite{nrp}: $l_1=20$, $l_2=10$, $\alpha=0.15$, $\epsilon=0.2$, and $\lambda=10$.

\noindent\textbf{PBG~\cite{lerer2019pbg}.}
We download the authors' official source codes\footnote{\url{https://github.com/facebookresearch/PyTorch-BigGraph}.}, and run the example script for Livejournal. For other datasets, we run the codes with default 30 iterations and report the best result.

\noindent\textbf{GraphVite~\cite{graphvite}.}
We adopt the reported results for YouTube, Friendster-small and Friendster in the original paper~\cite{graphvite}.  For other datasets, we run the authors' official source codes\footnote{\url{https://github.com/DeepGraphLearning/graphvite}.} with default setting. For Livejournal, other methods select $d=1024$, while the official implementation of GVT\footnote{ \url{https://graphvite.io/docs/0.2.1/api/solver.html##graphvite.solver.GraphSolver}} only allows the selection of $d$ up to $512$.

\noindent\textbf{NetSMF~\cite{netsmf}.}
We download the authors' official source codes\footnote{\url{https://github.com/xptree/NetSMF}.}, and run with the default hyper-parameter setting. 

\noindent\textbf{ProNE~\cite{prone}.}
We use the high-performance version of ProNE released by the LightNE GitHub Repository\footnote{\url{https://github.com/xptree/LightNE}.}, and keep the hyper-parameters the same as those in~\cite{prone}: $p=10, \theta=0.5, \mu=0.2$. 

\noindent\textbf{LightNE~\cite{lightne}.}
We download the authors' official source codes\footnote{\url{https://github.com/xptree/LightNE}.}, and run experiments with the default scripts and the parameter setting according to its original paper. 



\noindent\textbf{SketchNE.}
We set parameters $b=1, z=8, s_1=100, s_2=1000$ for all datasets and choose $T$ equal 2, 5 or 10, except on the very large graphs where we set $s_1=0$. 
We follow the embedding dimension $d$ setting in~\cite{lightne,graphvite,lerer2019pbg}, and let the other baselines follow the same setting. The eigen-decomposition rank $k$ should be larger than $d$. The parameters $\alpha$ and $q$ affect the experimental results to a larger extent. The bigger $q$, the more accurate the eigen-decomposition and thus the better the performance. We tune $\alpha$ and $q$ in the experiments, choosing $\alpha$ in the range \revise{$[0.35, 0.5]$} with step $0.05$ and $q$ in the range $[5,30]$ with step $5$. All parameter settings for SketchNE are listed in Table~\ref{tab:parameters}.
\begin{table}[h]
	\caption{Hyper-parameters for SketchNE.}
	\label{tab:parameters}
	\centering
	\small
	\renewcommand\arraystretch{0.9}
	\begin{tabular}{@{~}l@{~}|@{~}llllllll@{~}}
		\toprule
		Datasets     & $T$ & $k$ & $q$ & $\alpha$ & $s_1$ & $s_2$ & $d$  \\ \midrule
		BlogCatalog  & 10 &256 & 20 & 0.5 & 100 & 1000 & 128    \\ \midrule
		YouTube      & 10 &256 & 30 &  0.35 & 100 & 1000  & 128   \\ \midrule
		Friendster-small  &2 & 256 & 2  & 0.35 & 100& 1000 & 128    \\ \midrule
		Friendster      &2 & 128 & 2  &  0.4 & 100 & 1000 & 96   \\ \midrule
		OAG       & 10 & 256 & 30  &  0.45 & 100 & 1000  & 128   \\ \midrule
		Livejournal      & 5 & 1024 & 10  &  0.35 & 100 & 1000 & 1024   \\ \midrule
		ClueWeb       & 5 & 32 & 6  &  0.45 & 0 & 1000 & 32    \\ \midrule
		Hyperlink2014 & 5 &32 & 10  & 0.45 & 0 & 1000 & 32  \\ \midrule
		Hyperlink2012 & 5 &16 & 5 & 0.45 & 0 & 1000 & 16  \\ 
		\bottomrule 
	\end{tabular}
\end{table}

\revise{
\vpara{The Setting for GNN Methods.} 
To compare \model with GNN methods, we choose DGI~\cite{dgi}, GraphCL~\cite{you2020graphcl}, GCC~\cite{qiu2020gcc} and GraphSAGE~\cite{hamilton2017inductive} as the baselines. We download the official source codes released in the original papers, and run with the default hyper-parameter setting. 
Considering the fact that most GNNs require  vertex features as input, we follow the experimental setting in~\cite{yang2021iehgcn}, which is to generate random features for vertices. Specifically, for DGI and GraphCL, we generate a 128-dimensional random vector following the Xavier uniform or normal distribution~\cite{glorot2010understanding} for each vertex.
However, for GraphSAGE, we use its "identity features" as suggested in its original paper~(See Sec. 3.2 of \cite{hamilton2017inductive}) where each vertex has a learnable representation. 
DGI, GraphCL and GraphSAGE are trained on each dataset in an unsupervised way, we obtain the embedding matrix after the training process of these methods converges.
GCC is a pretrained GNN model which does not require training-from-scratch on downstream tasks, thus we download the pretrained model and then encode every vertex on our datasets to a 64-dimensional embedding. 
}

\vpara{The Setting for Vertex Classification.} 
To facilitate a fair comparison, we follow the training ratio setting in~\cite{deepwalk,graphvite,netsmf,lightne}.
A portion of labeled vertices are sampled for training and the remaining are used for testing. 
We complete the task by using one-vs-rest logistic regression implemented by LIBLINEAR~\cite{fan2008liblinear}. The prediction procedure is repeated five times and the average performance is evaluated in terms of both Micro-F1 and Macro-F1~\cite{tsoumakas2009mining}. 

\vpara{The Setting for Link Prediction.} For Livejournal, we follow the exactly same settings in Pytorch-BigGraph. 
For other three billion-scale graphs, we follow LightNE to set up the link prediction evaluation.
We randomly excludes 0.00001\% edges from the training graph for evaluation. 
When training SketchNE on these three graphs, the spectral propagation step is omitted due to memory cost and we set $d = 32$ except Hyperlink2012, where we use $d = 16$. 
We rank positive edges among randomly sampled corrupted edges to get the ranking metrics on the test set after training.
We evaluate the link prediction task with four metrics---mean rank (MR), HITS@10, HITS@50, and AUC.

\subsection{Experimental Results}
\label{exp:results}

\begin{figure*}[!htb]
	\centering
	\includegraphics[width=\linewidth]{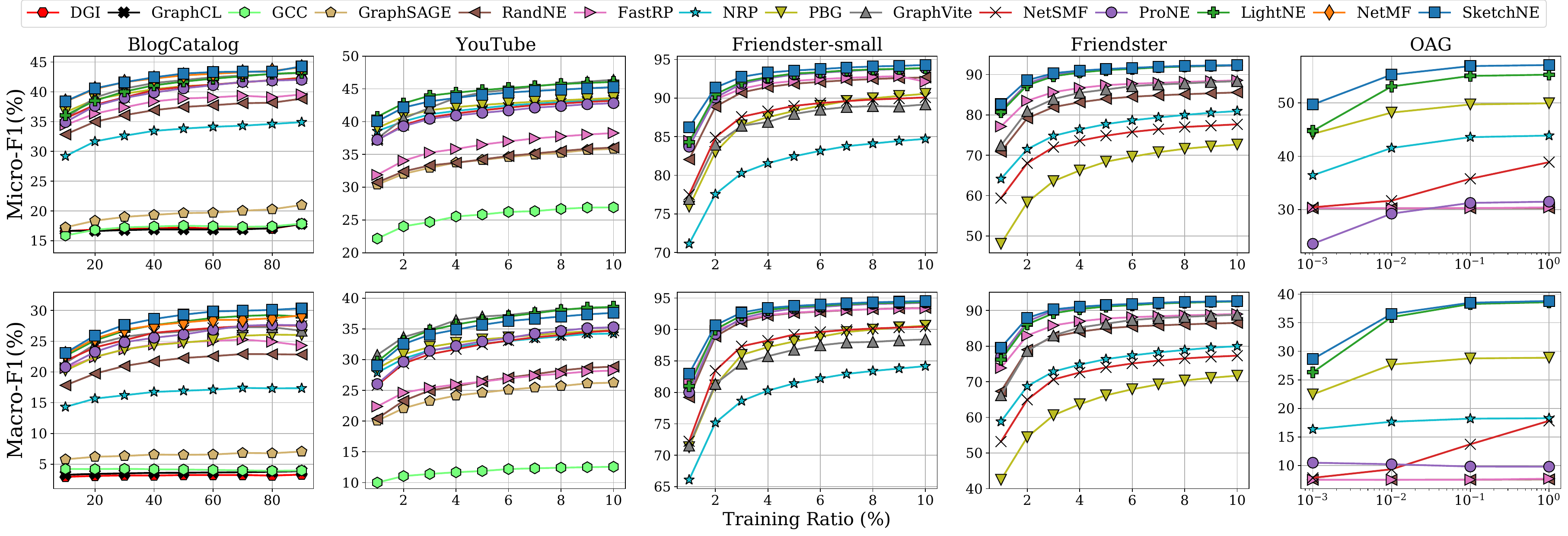}
	\vspace{-0.1in}
	\caption{\revise{Vertex classification performance~(Micro-F1 and Macro-F1) w.r.t. the ratio of training data. 
		\textmd{For methods that cannot handle computation or cannot finish job in one day, the results are not available and thus not plotted in this figure.}
	}}
	\vspace{-0.1in}
	\label{img2}
\end{figure*}

\vpara{Vertex Classification Results.}
We summarize the multi-label vertex classification performance in \Figref{img2}. 
In BlogCatalog, \model achieves significantly better Micro-F1 and Macro-F1 than the second best method LightNE (by 3.5\% on average). In YouTube~\cite{youtubedata}, SketchNE show comparable performance to LightNE and GraphVite, while show significantly better results than others. In OAG~\cite{oagdata}, SketchNE achieves better performance than LightNE---the second best baseline on this data (Micro-F1 improved by 5.4\% on average). In Friendster-small, and Friendster~\cite{friendsterdata}, SketchNE achieves the best performance among all baselines. 
\revise{
To illustrate the effectiveness of \model versus GNN methods, we test them on the BlogCatalog and YouTube datasets. The non-informative features with Xavier uniform distribution or Xavier normal distribution show almost the same performance and we retain the better results between the two distributions in \Figref{img2}. On BlogCatalog, \model achieves significantly better Micro-F1 and Macro-F1 than all of them---DGI, GraphCL, GCC, and GraphSage. 
On YouTube, GraphSAGE and GCC can complete the training, while GraphCL and DGI fail due to the limitation of GPU memory size. A recent work~\cite{zheng2022rethinking} revealed that DGI cannot scale to large graphs. Because GraphCL is built on top of DGI, it also cannot scale to large graphs. For very large graphs, all the GNN baselines cannot finish the training with its original code implementation due to either the limitation of GPU memory size or unconvergence within a reasonable time. 
Specifically, \model outperforms GraphSAGE on YouTube (28.9\% improvement for Micro-F1 and 44.6\% for Macro-F1 on average). \model achieves significantly better performance than GCC on YouTube (71.5\% improvement for Micro-F1 and 197.4\% for Macro-F1 on average). 

}

Overall, SketchNE has significantly better or comparable classification results compared to other methods. 
Compared to RandNE, FastRP and NRP, which omit element-wise function for scalability, 
SketchNE shows significantly better performance. 
It proves that the element-wise function is crucial for learning high quality embedding and that the method in \Secref{subsec:fLR} to factorize element-wise function of low rank matrix is practical. 
Overall, the vertex classification results illustrate the effectiveness superiority of \model.

\vpara{Link Prediction Results.}
Table~\ref{tab:linkpred} lists the link prediction performance.  
\revise{
For Livejournal, SketchNE outperforms all baselines in terms of MR, HITS@10, and HITS50. 
For the three billion-scale networks---ClueWeb, Hyperlink2014, and Hyperlink2012, we only report the results of LightNE and SketchNE, while other network embedding methods cannot finish running due to excessive memory or/and time cost, and GNN methods fail due to either the limitation of GPU memory size or unconvergence within a reasonable time.
}
The results of LightNE are reported by choosing edge sample parameters to reach 1.5 TB memory bound. 
On these datasets, \model produces significant outperformance over LightNE as measured by all four metrics. 
Take  Hyperlink2012---the largest one with 3.5 billion vertices and 225 billion edges---for example, \model achieves relative gains of 278\%, 282\%, 130\%, and 24\% over LightNE (the second best baseline on Livejournal) in terms of MR, HITS@10, HITS@50, and AUC.



 \begin{table}[!htb]
	\caption{Link prediction comparison. \revise{LJ, CW, HL14 and HL12 stands for Livejournal, ClueWeb, Hyperlink2014 and Hyperlink2012, respectively.}}
	\label{tab:linkpred}
	\small
	\renewcommand\arraystretch{0.9}
    \begin{tabular}{@{~}l@{~}|@{~}l@{~}|@{~~}l@{~~}l@{~~}l@{~~}l@{~}}
    \toprule
                   Datasets     & Systems & MR~$\downarrow$ & HITS@10~$\uparrow$ & HITS@50~$\uparrow$ & AUC~$\uparrow$ \\ \midrule
   \multirow{11}{*}{LJ} & \revise{GCC} & \revise{31.80}  &  \revise{0.335}  & \revise{0.735}   & \revise{0.689}\\
                    & \revise{GraphSAGE} & \revise{14.22}  &  \revise{0.655}  &  \revise{0.916}  & \revise{0.860}\\
                    & PBG & 4.25  &  0.929  & 0.974   & 0.968\\
                  & GraphVite & 3.06 & 0.962 & 0.982 & \textbf{0.973} \\
                   & NetSMF & 3.09 & 0.954  & 0.986   & 0.935 \\
                   & RandNE & 5.19 & 0.912  & 0.966   & 0.957 \\
                   & FastRP & 4.51 &  0.928  & 0.973   & 0.965 \\
                   & NRP & 2.98 & 0.949  & 0.993   & 0.934 \\
                   & ProNE & 3.31 & 0.950  & 0.982   & 0.932 \\
                   & LightNE & 2.13 & \textbf{0.977}  & 0.993   & 0.945 \\
                   & \textbf{\model} & \textbf{2.10} & \textbf{0.977}  & \textbf{0.994}   & 0.945
                   \\ \midrule
    \multirow{2}{*}{CW}       & LightNE & 105.9 & 0.753 & 0.803 & 0.903\\
                                   & \textbf{\model} & \textbf{32.0} &  \textbf{0.771} &  \textbf{0.869}& \textbf{0.968}    \\ \midrule
    \multirow{2}{*}{HL14} & LightNE  & 129.7 &  0.5 & 0.628  & 0.874\\
                                   & \textbf{\model} & \textbf{110.3} & \textbf{0.593}&\textbf{0.693}&\textbf{0.890} \\ \midrule
    \multirow{2}{*}{HL12} & LightNE & 257.7 & 0.189 & 0.348 & 0.751\\
                                   & \textbf{\model} & \textbf{68.1} & \textbf{0.722} &\textbf{0.802}  &\textbf{0.933}  \\ 
    \bottomrule 
    \end{tabular}
    \vspace{-0.1in}
\end{table}

\begin{table*}[t]
	\caption{Efficiency comparison among \model and other network embedding baselines.}
	\label{tab:runtimememory}
	\centering
	\small{
		\begin{threeparttable}
			\begin{tabular}{l@{}|@{}r@{~}|rrrrrrrr|r}
				\toprule
				Metric  &  Datasets & RandNE & FastRP & NRP & PBG &  GraphVite & NetSMF  & ProNE & LightNE & \textbf{\model} \\ \midrule
				\multirow{9}{*}{Time} & BlogCatalog & \textbf{0.5~s} & 1.0~s & 3.0~s & 174.0~s & 4.0~s & 11.3~m  & 79.0~s & 152.0~s & 2.0~s       \\
				& YouTube    & \textbf{12.0~s}  & 17.0~s &  173.0~s & 12.5~m & 44.0~s & 3.7~h & 65.0~s & 96.0~s  & 40.0~s         \\
				& Friendster-small &11.8~m & 40.0~m & 3.5~h  & 22.7~m & 2.8~h & 52~m & 5.3~m  & 7.5~m   & \textbf{5.2~m}         \\
				& Friendster & 57.8~m &3.5~h& 16.1~h   & 5.3~h &  20.3~h & 16.5~h & 19.5~m & 37.6~m  & \textbf{16.0~m}        \\
				& OAG  & 33.7~m & 3.5~h &  11.4~h  & 20~h & 1+day & 22.4~h &\textbf{22.6~m}  & 1.5~h  & 1.1~h        \\
				& Livejournal & \textbf{9.0~m} & 18.0~m & 4.3~h  & 7.3~h & 29.0~m & 2.1~h & 12.8~m  & 16.0~m  & 12.5~m \\
				& ClueWeb &$\times$ &$\times$ & $\times$  & 1+day &  1+day &$\times$ &$\times$  & 1.3~h & \textbf{37.7~m}        \\
				& Hyperlink2014 &$\times$ &$\times$ & $\times$  &1+day & 1+day &$\times$ & $\times$ & 1.8~h\tnote{1} & \textbf{0.98~h}        \\
				& Hyperlink2012 &$\times$ &$\times$ &$\times$ &1+day & 1+day &$\times$ & $\times$ & 5.6~h\tnote{1} & \textbf{1.0~h}        \\ \midrule
				\multirow{8}{*}{\makecell{Mem\\(GB)}}& BlogCatalog &\textbf{0.2} & 0.3 &0.6  &$\star$  &   $\star$  & 135  & 18  & 273   & 17       \\
				& YouTube  &\textbf{6.9} & 12.5 &9.7  &$\star$  & $\star$ & 854 &   28  & 83    & 27        \\
				& Friendster-small &125 &125 &105 & $\star$ & $\star$ & 85 & 84  & 541   & \textbf{56}        \\
				& Friendster & 548& 583 & 400 &$\star$  & $\star$ & 1144 &  326   & 559  & \textbf{236}         \\
				& OAG   &429 &746 & 473   & $\star$ & $\star$ &1500 & 403  & 1391  & \textbf{283}       \\
				& Livejournal &224 & 417 &282  & $\star$ & $\star$ & 140 & \textbf{131}  & 532 &147 \\
				& ClueWeb  &$\times$ &$\times$ & $\times$ & $\star$ & $\star$  &$\times$ &$\times$ & 1493  & \textbf{612}    \\
				& Hyperlink2014 &$\times$ &$\times$ &$\times$  & $\star$ & $\star$ &$\times$ &   $\times$ & 1500\tnote{1}  & \textbf{1076}         \\
				& Hyperlink2012 & $\times$&$\times$ &$\times$  & $\star$ & $\star$ & $\times$ & $\times$ & 1500\tnote{1}  & \textbf{1321}       \\ \bottomrule
			\end{tabular}
			\begin{tablenotes}
				\footnotesize
				\item[1] LightNE requires sufficient samples that cost more than 1500GB mem (and more time), so it has to stop before it reaches the mem limit.
				\\"$\star$" indicates that we do not compare the memory cost of the CPU-GPU hybrid system (GraphVite) or distributed memory system (PBG). 
				\\"$\times$" indicates that the corresponding algorithm is unable to handle the computation due to excessive space and memory consumption.
			\end{tablenotes}
		\end{threeparttable}
	}
\end{table*}

\revise{
\vpara{More Discussion on the Performance of GNNs.} 
It is interesting to see that network embedding~(NE) methods (not only our \model, but also other NE methods such as FastRP, NRP and ProNE) significantly outperform GNN methods in vertex classification on BlogCatalog and YouTube datasets, as well as link prediction on Livejournal dataset.
We attribute the poor performance of GNNs  to the nature of the three datasets themselves. These datasets (and corresponding tasks) are mainly proximity-based, so the inductive bias of NE methods  that  proximal vertices should have similar representations can significantly boost their performance. However, GNN methods usually consider more complex and general information such as structure and attributes, which may limit their performance in proximity-based tasks.
}

\revise{
\begin{table}[h]
	\caption{\revise{Time comparison among \model and GNN baselines. "×" indicates that the corresponding algorithm is unable to handle the computation due to the limited-size GPU memory. Larger datasets are not listed due to the scalability issue of GNN models.}}
	\label{tab:time4gnns}
	\centering
	{\color{black}\begin{tabular}{@{~}r@{~}|@{~}r@{~}|@{~}r@{~}|@{~}r@{~}}
		\toprule
		Baselines & BlogCatalog & YouTube & Livejournal \\
		\midrule
		DGI & 6.3 m & $\times$ & $\times$ \\
		GraphCL & 15.5 m & $\times$ & $\times$\\
		GCC & 10.9 m &  0.98 h& 15.6 h \\
		GraphSAGE & 6.2 h& 18.0 h & 46.5 h\\\midrule
		\model & \textbf{2.0 s}& \textbf{40.0 s} & \textbf{12.5 m}\\
		\bottomrule
	\end{tabular}}
\end{table}
}

\vpara{Time and Memory Efficiency.}
We report the running time and memory cost of \model and other eight network embedding baselines on all nine datasets in 
Table~\ref{tab:runtimememory}. 
Time-wise, on small datasets with millions of edges, the running time of \model is relatively comparable to other baselines. 
However, on networks of billions of edges, e.g., Friendster, ClueWeb, Hyperlink2014 and Hyperlink2012, it takes \model the least time to embed them. \revise{
The GNN baselines, DGI and GraphCL are limited by GPU memory for the full-graph training and difficult to scale to large graphs, while GCC and GraphSAGE show slow convergence because it is based on graph sampling. For example, the training of GraphSAGE (trained via neighborhood sampling, a popular graph sampling method) on BlogCatalog (the smallest graph which consists of 10,312 vertices and 333,983 edges) using a GeForce GTX 1080 Ti GPU costs more than six hour, while it costs only 2 seconds for SketchNE to run the same task. We list all the time comparison among \model and GNN baselines in Table~\ref{tab:time4gnns}.
}
Memory-wise, we can observe that \model demands less memory than all baselines on all datasets except using slightly more memory than NetSMF and ProNE on the small Livejournal data, empowering it to go for the largest networks considered and beyond. 
For example, the running time of SketchNE on ClueWeb~\cite{cluewebdata} is 37.7 minutes and the peak memory cost is 612GB, 
which is $2\times$ faster than LightNE and saves more than 59\% memory. 
The results on Hyperlink2014 and Hyperlink2012~\cite{hyperlinkdata} further demonstrate the efficiency of SketchNE. It is worth noting that SketchNE can embed the Hyperlink2012 network with 3.5 billion  vertices and 225 billion edges in 1.0 hours by using 1,321GB memory on a single machine. 
In conclusion, the results on the three very large-scale graphs demonstrate that SketchNE can achieve consistently and significantly better efficiency than LightNE in terms of both running time and memory cost.  


\subsection{Ablation and Case Studies} 
\label{exp:ablation}
 \begin{figure}[!htb]
     \centering
     \subfloat[The computed eigenvalues w.r.t. $\alpha$]{
         \includegraphics[width=0.49\linewidth]{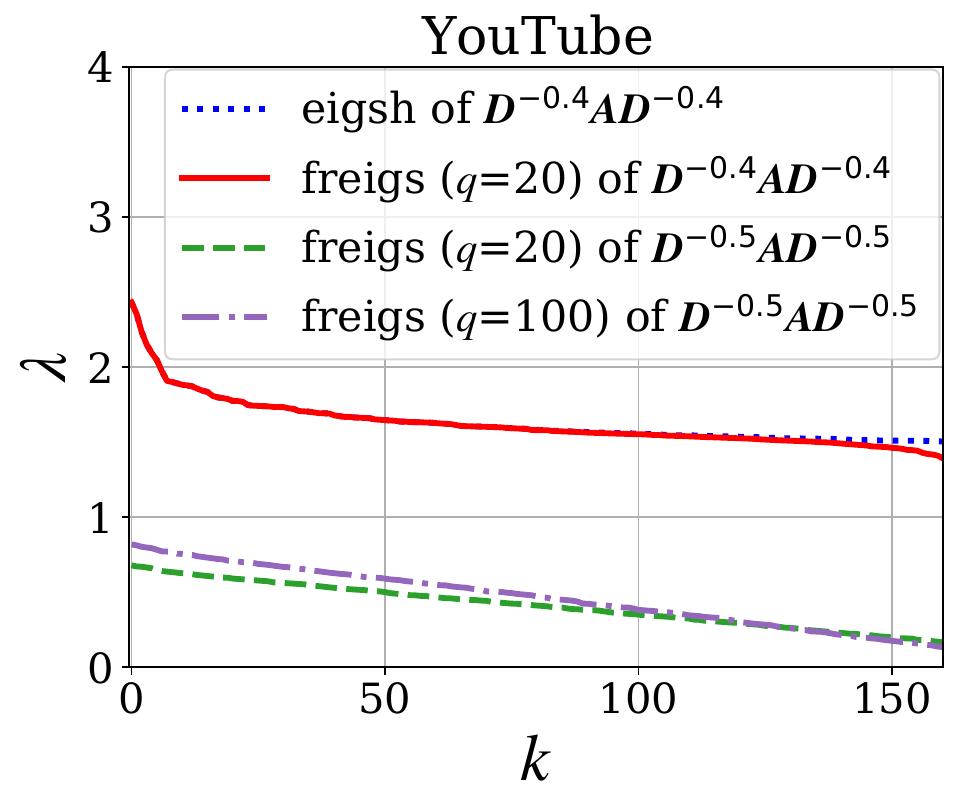}
     }
     \subfloat[ NetMF vs. NetMF (w/ freigs)]{
         \includegraphics[width=0.49\linewidth]{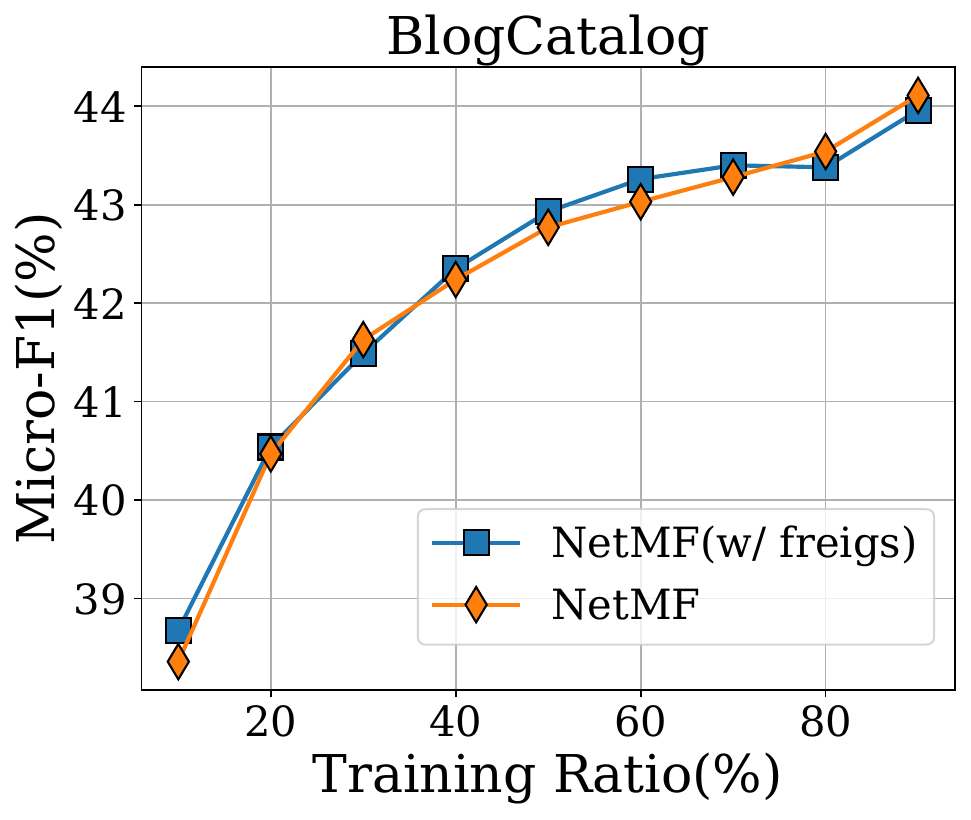}
     }
     \caption{
        The validation of the effectiveness of freigs.
     }
     \label{fig:freigs}
 \end{figure}

\begin{figure}[t!]
	\centering
	\includegraphics[width=\linewidth]{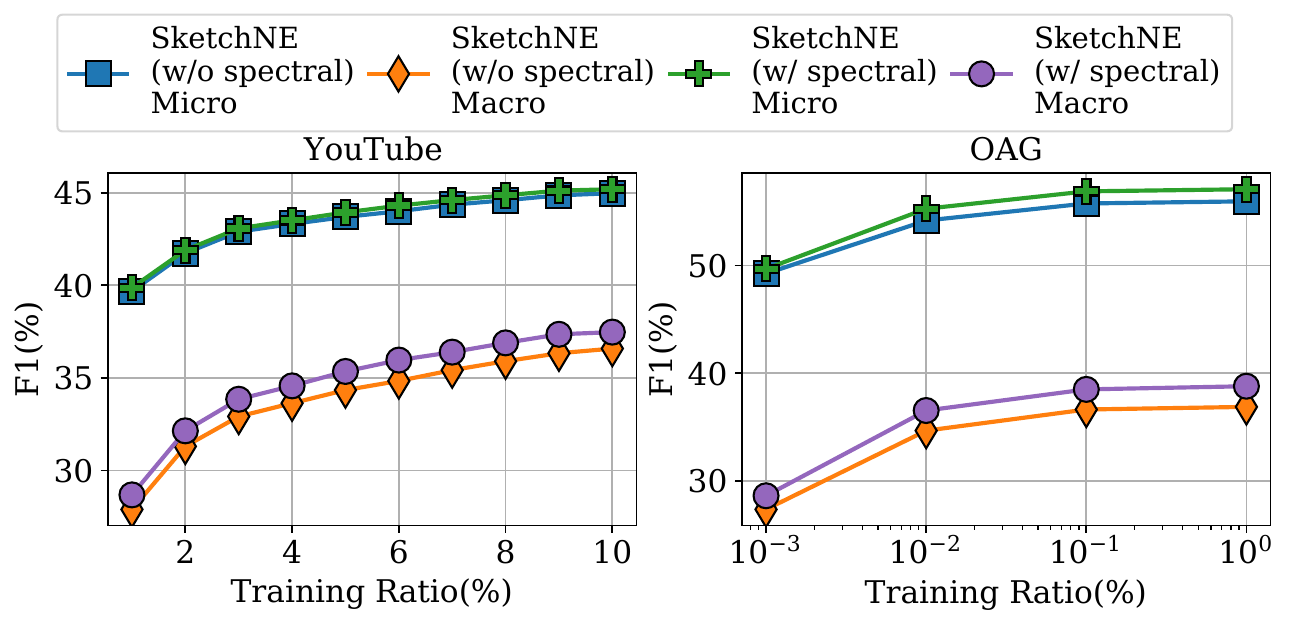}
	\caption{The embedding performance comparison.}
	\label{fig:comparespectral}
\end{figure}

\vpara{Efficiency and Effectiveness of Each Step of \model.} 
First, we focus on the fast randomized eigen-decomposition (\texttt{freigs}, \Algref{alg:SketchNE} Line 1). Halko et al.~\cite{halko2011finding} has shown that it is challenging to perform eigen-decomposition on $\vect{D}^{-1/2}\vect{A}\vect{D}^{-1/2}$. 
We choose YouTube as an example, whose related eigenvalues are shown in \Figref{fig:freigs}~(a). 
The eigenvalues of $\vect{D}^{-1/2}\vect{A}\vect{D}^{-1/2}$ with power iteration $q=20$ and $q=100$ are far from correct, as  the correct largest eigenvalue should be 1. 
The \texttt{eigsh}~\cite{arpack1998} is not able to complete this job in three days. The results illustrate the requirement to use modified Laplacian matrix  for fast randomized eigen-decomposition.
When we choose $\alpha = 0.4$, the eigenvalues of fast randomized eigen-decomposition ($q=20$) on $\vect{D}^{-0.4}\vect{A}\vect{D}^{-0.4}$ is indistinguishable from the eigenvalues computed by \texttt{eigsh} when $k<140$. The running time of fast randomized eigen-decomposition is 20 seconds while \texttt{eigsh} costs 31 minutes, which proves the accuracy and efficiency of fast randomized eigen-decomposition. Then, we replace the eigen-decomposition~(\texttt{eigsh}) of NetMF with \Algref{alg:frevd} and evaluate the Micro-F1 result of BlogCatalog between NetMF and NetMF~(w/ freigs) in \Figref{fig:freigs}~(b). The results show the effectiveness of \texttt{freigs}.
Next, we focus on the effects of sparse-sign randomized single-pass SVD and spectral propagation.  \model~(w/o spectral) and \model~(w/ spectral) represent the result of initial and enhanced embeddings, respectively. 
The vertex classification result is shown in \Figref{fig:comparespectral}.
The performance of model~(w/o spectral) has been satisfactory, which proves the effectiveness of the sparse-sign randomized single-pass SVD. Combining with spectral propagation, \model~(w/ spectral) shows better results, 
demonstrating the effect of spectral propagation. 

\begin{figure}[!t]
	\centering
	\includegraphics[width=\linewidth]{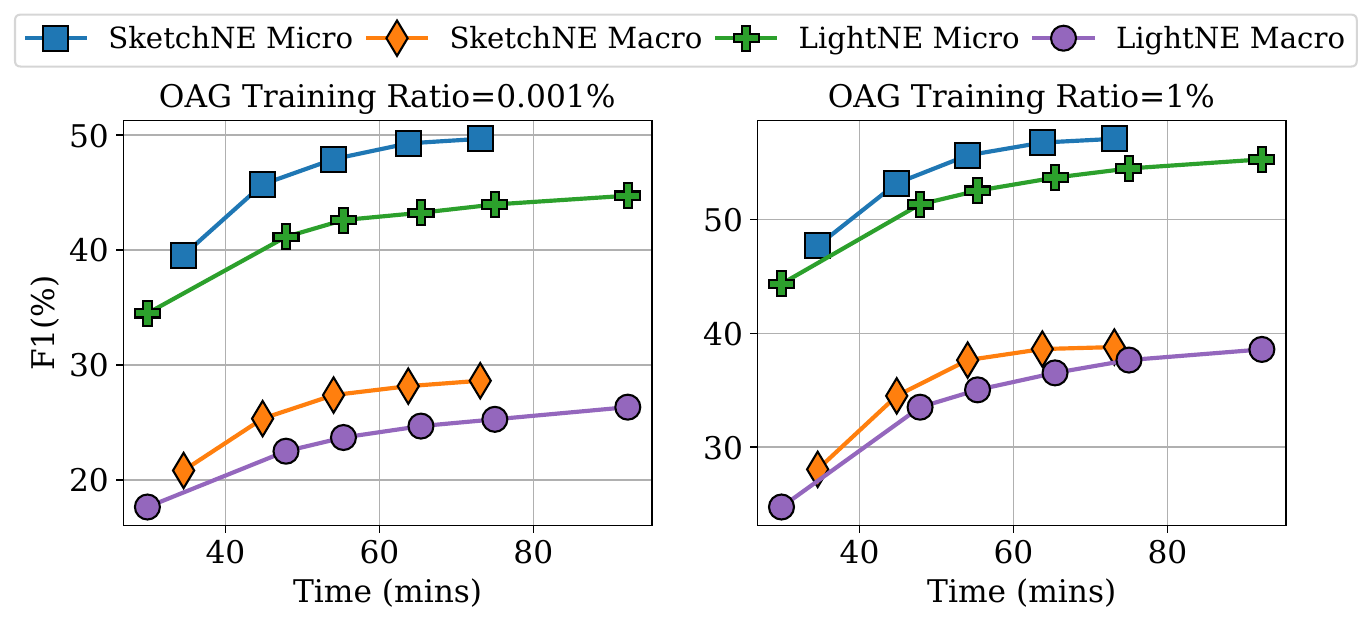}
	\caption{The trade-offs between efficiency and performance.}
	\label{fig:oagee}
\end{figure}

\vpara{\revise{The Effects of Parameter $q$.}}
Here we need to pay attention to $q$, which determines the accuracy of the fast randomized eigen-decomposition. Thus, we make a trade-off between the quality of the learned embeddings and the overall running time. The peak memory cost is constant when we fix the oversampling parameters, column density $z$, eigen-decomposition rank $k$, and embedding dimension $d$. 
By fixing other parameters for OAG, we enumerate  $q$ from $\{10,15,20,25,30\}$, the efficiency-effectiveness trade-off of OAG is shown as \Figref{fig:oagee}. 
We also add LightNE results with different edge samples $M$ from $\{1Tm,5Tm,7Tm,10Tm,13Tm,17Tm\}$. 
The peak memory of SketchNE is still {283GB} while that of LightNE is {553 GB, 682 GB, 776 GB, 936 GB, 1118 GB} and \textbf{1391 GB}, respectively. \Figref{fig:oagee} shows that SketchNE can learn more effective embeddings than LightNE when the running time is constant with less memory cost. The experiment proves that users can adjust \model flexibly according to time/memory budgets and performance requirements.

\revise{
\vpara{The Effects of Parameter $\alpha$.}
We also analyze the influence of parameter $\alpha$, which balances the accuracy of fast randomized eigen-decomposition on modified Laplacian matrix and the approximation error of Eq. (\ref{eq11}). We select here YouTube and OAG as example datasets for the ablation study. We vary the parameter $\alpha$ from $\{0.25,0.3,0.35,0.4,0.45,0.5\}$ and fix other parameters. The change of \model's performance as the $\alpha$ varies can be shown in \Figref{fig:alpha}. From it we can see that when setting $\alpha=0.5$, the eigen-decomposition on the Laplacian matrix has a bad accuracy and therefore causes a loss of performance in \Figref{fig:alpha}. \Figref{fig:alpha} also shows that \model can learn almost the best embeddings when setting $\alpha$ in the range $[0.35, 0.45]$.
\begin{figure}[!t]
	\centering
	\includegraphics[width=\linewidth]{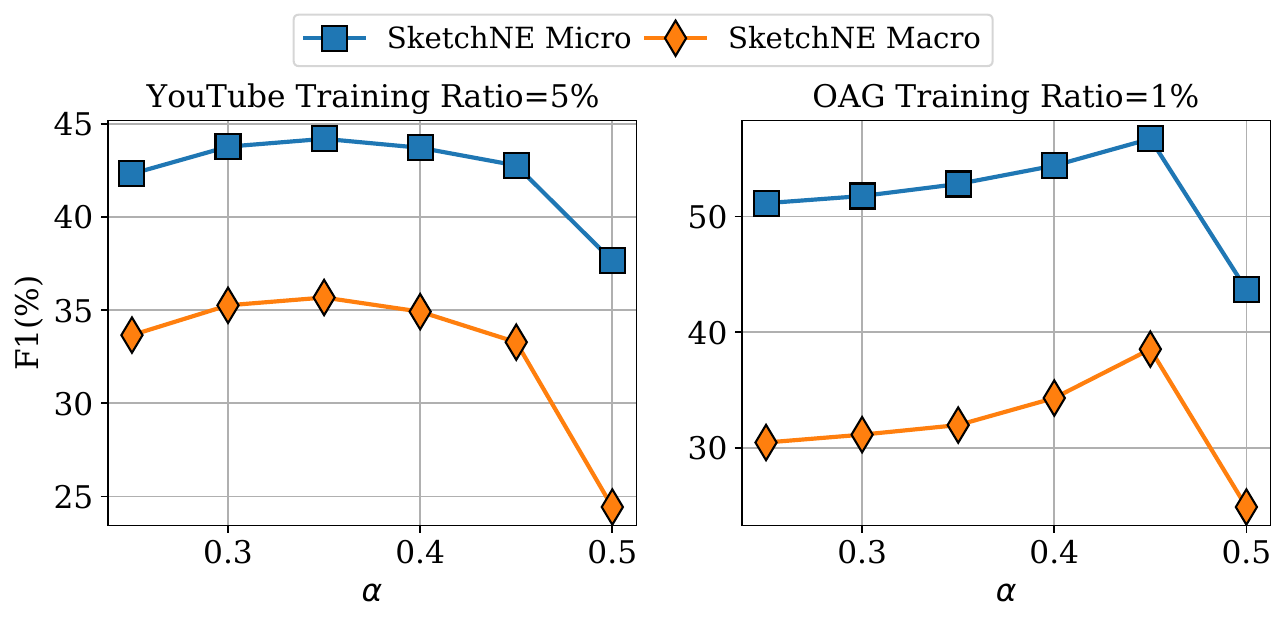}
	\caption{\revise{The performance change when varying parameter $\alpha$.}}
	\label{fig:alpha}
\end{figure}

}
\vpara{The Sketching Method.}
We download the official source codes\footnote{\url{https://github.com/insuhan/polytensorsketch}.} of Polynomial Tensor Sketch~\cite{han2020polynomial}, and set $k=10, r=11, m=12$, which makes dimension $m\times r+1$ slightly bigger than SketchNE's $d=128$ for BlogCatalog.
We replace the sparse-sign randomized single-pass SVD in \model
with the polynomial tensor sketch algorithm, which is an alternative solution to factorize $f^\circ(\mL \mR)$. 
Experiments on BlogCatalog show that SketchNE with sparse-sign randomized single-pass SVD performs much better than 
SketchNE with polynomial tensor sketch, achieving a 44\%/67.8\% relative improvement on Micro-F1/Macro-F1 with 10\% of training data.

\begin{figure}[!t]
	\centering
	\includegraphics[width=0.85\linewidth]{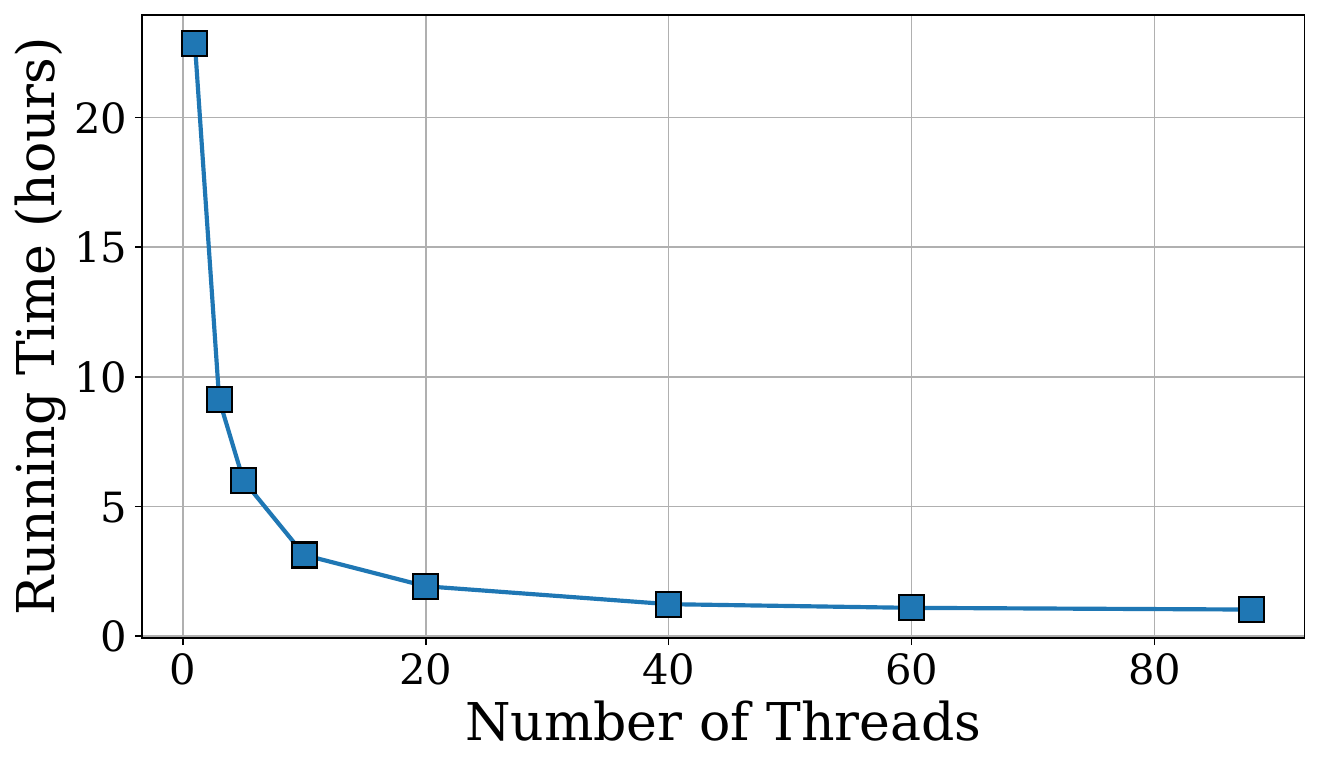}
	\caption{Runtime of SketchNE v.s. the number of threads.}
	\label{threads}
\end{figure}

\vpara{The Number of Threads.}
In this work, we use a single-machine shared memory implementation with multi-threading acceleration.
We set the number of threads to be 1, 3, 5, 10, 20, 40, 60, 88, and report the corresponding running time of \model in \Figref{threads}.
\model takes 22.8 hours to embed the Hyperlink2012 network with 1 threads and 1.9 hours with 40 threads, achieving a 12$\times$ speedup 
ratio~(with ideal being 40$\times$). This relatively good sub-linear speedup
supports \model to scale up to networks with hundreds of billions of edges.

\section{Related Work}
\revise{In this section, we first review the related work of network embedding and graph neural networks, and dicuss their difference from a vertex similarity perspective.
Then we review related work on randomized matrix factorization.}

\vpara{Network Embedding~(NE).} Network embedding has been comprehensively studied over the past decade. Recent work about network embedding can be divided into two categories. The first category is based on skip-gram methods inspired by word2vec~\cite{word2vec}, including DeepWalk~\cite{deepwalk}, Line~\cite{line}, node2vec~\cite{node2vec}. These methods rely on stochastic gradient descent to optimize a logistic loss. 
The second category method is based on matrix factorization, using SVD or other matrix decomposition techniques to generate the best low-rank approximation~\cite{eckart1936approximation}. GraRep~\cite{grarep}, HOPE~\cite{hope}, NetMF~\cite{netmf}, NetSMF~\cite{netsmf}, ProNE~\cite{prone}, NRP~\cite{nrp}, LightNE~\cite{lightne} and PANE~\cite{yang2020pane} are methods in this category. There are several high performance embedding systems for large graphs have been developed. GraphVite~\cite{graphvite}, a CPU-GPU hybrid network embedding system, is developed based on DeepWalk~\cite{deepwalk} and LINE~\cite{line}. In GraphVite, CPU is used to perform graph operation, and GPU is used to compute linear algebra. Nevertheless, the GPU memory is a disadvantage when processing billion scale networks, limiting widespread use. Based on DeepWalk~\cite{deepwalk} and LINE~\cite{line}, PyTorch-BigGraph~\cite{lerer2019pbg} has been proposed for distributed memory machines. It achieves load balancing by graph partition and synchronization through a shared parameter server. In this work, we propose SketchNE, which leverages the merit of NetSMF and LightNE, and addresses their limitation in speed and memory overhead. 

\revise{\vpara{Graph Neural Networks~(GNNs).} GNNs introduce deep neural networks into graph learning, including GCN~\cite{gcn}, GAT~\cite{gat}, GIN~\cite{gin}, ie-HGCN~\cite{yang2021iehgcn}, GCC~\cite{qiu2020gcc}, GraphCL~\cite{you2020graphcl}, DGI~\cite{dgi} and GraphSAGE~\cite{hamilton2017inductive}. 
In practice, scaling GNN models to large graphs can be time consuming. The training process of GNN models can be classified into two main 
types. The first type is full-graph training where each gradient descent step requires traversing the full graph and thus the time complexity is proportional to the graph size. The second type is graph sampling-based where each gradient descent step only involves sampled subgraphs. Although the time complexity for each gradient descent step is significantly reduced by sampling, the number of gradient descent steps required for convergence is empirically proportional to the graph size, which makes the overall time complexity proportional to the graph size. Therefore, training GNNs on billion-scale graphs relies on distributed computing supports with both CPUs and GPUs, such as the one used for the PinSage~\cite{pinsage}, Psgraph~\cite{jiang2020psgraph}, AGL~\cite{zhang2020agl} and Neugraph~\cite{ma2019neugraph} systems, while the premise of SketchNE is to embed billion-scale graphs into latent embeddings with only a CPU server efficiently.

In this work, we also compare several GNNs with our proposed network embedding solution, \model.
For a fair comparison with network embedding methods which mainly conduct unsupervised learning on graphs without attributes, we focus on GNNs which can be trained (1) in an unsupervised/self-supervised way and (2) without additional vertex and edge features, including GraphSAGE~\cite{hamilton2017inductive}, DGI~\cite{dgi}, GCC~\cite{qiu2020gcc}, GraphCL~\cite{you2020graphcl}, GraphMAE~\cite{hou22graphmae}, and ie-HGCN~\cite{yang2021iehgcn}. GraphSAGE and DGI propose loss functions which can be used to train GNNs models in an unsupervised manner. GCC proposes a framework to capture the universal network topological properties in a self-supervised way. GraphCL improves the performance of graph self-supervised learning by data augmentations. ie-HGCN replaces vertex attributes in GNNs with random initialized features.
}

\vpara{Randomized Matrix Factorization.} As the amount of data continues to increase, the popularity of randomized matrix computations has grown significantly. Randomized SVD can be an alternative to conventional SVD methods, because it involves the same or fewer floating-point operations and is more efficient for truly large high-dimensional data, by exploiting modern computing architectures~\cite{halko2011finding,yu2018efficient,persvd}. As an application, frPCA~\cite{frpca} is developed for large-scale sparse data with better performance than conventional PCA algorithms. Randomized SVD has shown superiority in recent network embedding methods such as NetSMF, ProNE, and LightNE. Over the past few years, several single-pass SVD algorithms~\cite{yu2017ijcai,tropp2019streaming} based on randomized matrix sketch have been introduced for streaming data scenarios. The efficiency and memory limitation of NetMF will be solved by randomized matrix factorization.

\section{Conclusion}

In this work, we propose \model, a fast, memory-efficient, and scalable network embedding method. 
We formulate the computation goal of NetMF as 
factorizing an element-wise function of low-rank matrix and then analyze its computational challenges.
\model resolves these challenges by leveraging various randomized linear sketch techniques,
including but not limited to sparse-sign matrix, single-pass SVD, and fast eigen-decomposition. We use sparse-sign random projection matrix to solve the matrix multiplication challenge between $f^{\circ}(\vect{L}\vect{R})$ and random projection matrix, and generate a sketch that can caputure the dominant information of $f^{\circ}(\vect{L}\vect{R})$. 
Then, we solve the challenge in constructing reduced matrix by single-pass SVD. Secondly, we propose a fast randomized eigen-decomposition algorithm for modified Laplacian matrix. 
To enhance the performance of embedding, spectral propagation is adopted and a high-performance parallel graph processing stack GBBS is used to achieve memory-efficiency.  
The main computation steps of SketchNE are highly parallelizable, which is thus well supported by the MKL library and OpenMP. 
With the help of these techniques, \model achieves the best performance on vertex classification and link prediction among state-of-the-art methods across diverse datasets. 
Notably, \model can learn high-quality embeddings for a network with 3.5 billion vertices and 225 billion edges in 1.0 hour by using 1,321GB memory on a single machine, and the learned embeddings offer a \textbf{282\%} relative HITS@10 improvement over LightNE on the link prediction task.  

In the future, we plan to extend the method to handle dynamic and heterogeneous networks.
Since the sparse-sign randomized single-pass SVD is proposed for solving the problem $f^{\circ}(\vect{L}\vect{R})$, it is foreseeable that \Algref{alg:ssrspSVD} may have advantages in approximating the basic compoents in deep learning network. 
\ifCLASSOPTIONcaptionsoff
  \newpage
\fi



\bibliographystyle{IEEEtran}
\bibliography{IEEEabrv}
%






\end{document}


\appendices

\section{Theoretical Proof}
The following lemmas will be useful in our proof.
\setcounter{lemma}{0}
\renewcommand{\thelemma}{\textbf{\arabic{lemma}}}
\setcounter{theorem}{0}
\renewcommand{\thelemma}{\textbf{\arabic{lemma}}}
\setcounter{equation}{0}
\renewcommand{\theequation}{A.\arabic{equation}}

\begin{lemma}
	\label{lemma1}
	(\cite{trefethen1997numerical}) Singular values of a real symmetric matrix are the absolute values of its eigenvalues.
\end{lemma}

\begin{lemma}	
	\label{lemma2}
	(\cite{halko2011finding,musco2015nips}) For input matrix $\vect{X}$, the Alg.~2 with the power iteration scheme has the following guarantee: 
	\beq{\nonumber
		\Vert\vect{X}-\vect{Q}\vect{Q}^{\top}\vect{X}\Vert_F=\Vert\vect{X}-\vect{U}\vect{\Sigma}\vect{V}^{\top}\Vert_F\leq (1+\varepsilon)\Vert\vect{X}-\vect{X}_k\Vert_F,
	}with high probability. $\vect{X}_k$ is the best rank-$k$ approximation of $\vect{X}$.
\end{lemma}

\begin{lemma}	
	\label{lemma3}
	(\cite{horn1994topics}) Let $\vect{B},\vect{C}$ be two $n \times n$ symmetric matrices. Then for the decreasingly ordered singular values $\sigma$ of $\vect{B},\vect{C}$ and $\vect{BC}$, $\sigma_{i+j-1}(\vect{BC})\leq \sigma_i(\vect{B})\times \sigma_j(\vect{C})$ holds for any $1\leq i,j\leq n$ and $i+j\leq n+1$.
\end{lemma}

\begin{lemma}	
	\label{lemma4}
Suppose $\vect{U}_k\vect{\Lambda}_k\vect{U}_k^{\top}$ is the rank-$k$ truncated eigen-decomposition of $\vect{D}^{-\alpha}\vect{A}\vect{D}^{-\alpha}$, obtained with Alg. 5.
Then, \\
	$\Vert\vect{D}^{-\frac{1}{2}}\vect{A}\vect{D}^{-\frac{1}{2}}-\vect{D}^{-\frac{1}{2}+\alpha}\vect{U}_k\vect{\Lambda}_k\vect{U}_k^{\top}\vect{D}^{-\frac{1}{2}+\alpha}\Vert_F\leq(1+\varepsilon)d_{min}^{-1+2\alpha}\sqrt{\sum_{j=k+1}^{n}|\lambda_j|^2}$
	with high probability. Here, $|\lambda_j|$ is the $j$-th largest absolute value of eigenvalue of $\vect{D}^{-\alpha}\vect{A}\vect{D}^{-\alpha}$ and $d_{min}$ is the minimum vertex degree.
\end{lemma}


\begin{proof}
	Applying Lemma~\ref{lemma1} and Lemma~\ref{lemma2}, we see that the results of  Alg.~5 satisfy
	\beq{
		\label{proof_eq1}
		\besp{
			&\Vert\vect{D}^{-\alpha}\vect{A}\vect{D}^{-\alpha}\!-\!\vect{U}_k\vect{\Lambda}_k\vect{U}_k^{\top}\Vert_F\\
			&\!\leq \!(1+\varepsilon) \Vert\vect{D}^{-\alpha}\vect{A}\vect{D}^{-\alpha}\!-\!(\vect{D}^{-\alpha}\vect{A}\vect{D}^{-\alpha})_k\Vert_F\\
			&\!=\!(1+\varepsilon)\sqrt{\sum_{j=k+1}^{n}|\lambda_j|^2} ~,
	}}where $|\lambda_j|$ is the $j$-th largest absolute value of eigenvalue of $\vect{D}^{-\alpha}\vect{A}\vect{D}^{-\alpha}$. 
Notice that Alg. 5 is mathematically equivalent to Alg. 2 in exact arithmetic.
Suppose $\sigma_i(\cdot)$	denotes the $i$-th singular value of a matrix. Then,
	\beq{
		\label{proof_eq2}
		\besp{
			&\left\Vert\vect{D}^{-1/2}\vect{A}\vect{D}^{-1/2}-\vect{D}^{-1/2+\alpha}\vect{U}_k\vect{\Lambda}_k\vect{U}_k^{\top}\vect{D}^{-1/2+\alpha}\right\Vert_F=\\
			&\sqrt{\sum_{i=1}^{n}\sigma_i\left(\vect{D}^{-1/2}\vect{A}\vect{D}^{-1/2}-\vect{D}^{-1/2+\alpha}\vect{U}_k\vect{\Lambda}_k\vect{U}_k^{\top}\vect{D}^{-1/2+\alpha}\right)^2}.
	}}Here, we have $\alpha\in(0,0.5]$. Notice that the smallest vertex degree $d_{min}$ satisfies that $d_{min}^{-1/2+\alpha}$ is largest singular value of $\vect{D}^{-1/2+\alpha}$
	Applying Lemma~\ref{lemma3} twice  to (\ref{proof_eq2}), we  have
	\beq{
		\label{proof_eq3}
		\besp{
			&\Vert\vect{D}^{-1/2}\vect{A}\vect{D}^{-1/2}-\vect{D}^{-1/2+\alpha}\vect{U}_k\vect{\Lambda}_k\vect{U}_k^{\top}\vect{D}^{-1/2+\alpha}\Vert_F\leq\\
			&\sqrt{\sum_{i=1}^{n}\sigma_{1}(\vect{D}^{-1/2+\alpha})^2\sigma_{i}(\vect{D}^{-\alpha}\vect{A}\vect{D}^{-\alpha}-\vect{U}_k\vect{\Lambda}_k\vect{U}_k^{\top})^2 \sigma_{1}(\vect{D}^{-1/2+\alpha})^2}\\
			&=d_{min}^{-1+2\alpha}\sqrt{\sum_{i=1}^{n}\sigma_{i}(\vect{D}^{-\alpha}\vect{A}\vect{D}^{-\alpha}-\vect{U}_k\vect{\Lambda}_k\vect{U}_k^{\top})^2}.
	}}
	Combining (\ref{proof_eq1})  and (\ref{proof_eq3}), we derive
	\beq{
		\nonumber
		\besp{
			&\Vert\vect{D}^{-1/2}\vect{A}\vect{D}^{-1/2}-\vect{D}^{-1/2+\alpha}\vect{U}_k\vect{\Lambda}_k\vect{U}_k^{\top}\vect{D}^{-1/2+\alpha}\Vert_F\\
			&\leq d_{min}^{-1+2\alpha}\sqrt{\sum_{i=1}^{n}\sigma_{i}(\vect{D}^{-\alpha}\vect{A}\vect{D}^{-\alpha}\!-\!\vect{U}_k\vect{\Lambda}_k\vect{U}_k^{\top})^2}
			\!\\
			&\leq\!(1+\varepsilon)d_{min}^{-1+2\alpha}\sqrt{\sum_{j=k+1}^{n}|\lambda_j|^2}.
		}
	}
	Then, Lemma~\ref{lemma4} is proved.
\end{proof}

\begin{theorem}
\label{thm:thm1}
Suppose $f^{\circ}$ denotes $\mathrm{trunc\_log}^{\circ}$, i.e. the element-wise truncated logarithm, $f^{\circ}(\vect{M})$ is the matrix in (1), and $f^{\circ}(\vect{L^{\prime}R^{\prime}})$ is defined by (11) which includes the quantities obtained with Alg. 5. Then,\\
	\begin{equation*}
        \Vert f^{\circ}(\vect{M})\!-\!f^{\circ}(\vect{L^{\prime}R^{\prime}})\Vert_F\!\leq\!\frac{(1+\varepsilon)d_{min}^{-1+2\alpha}B}{(c-1)bT}
    \end{equation*}
	with high probability. Here $|\lambda_j|$ is the $j$-th largest absolute value of eigenvalue of $\vect{D}^{-\alpha}\vect{A}\vect{D}^{-\alpha}$, $B=\sqrt{\sum_{j=k+1}^{n}|\lambda_j|^2}((c^T\!-\!1)(1+\frac{n}{c\!-\!1})\!-\!nT)\mathrm{vol}(G)$, $c=n(\frac{d_{max}}{d_{min}})^{1-2\alpha}$. $d_{min}$ and $d_{max}$ are the minimum and maximum vertex degrees, respectively.
\end{theorem}

\begin{proof}
	Suppose $\vect{Y} = \vect{D}^{-1/2+\alpha}\vect{U}_k\vect{\Lambda}_k\vect{U}_k^{\top}\vect{D}^{-1/2+\alpha}$ and $\vect{X} = \vect{D}^{-1/2}\vect{A}\vect{D}^{-1/2}$.
	With $\vect{M} = \frac{\mathrm{vol}(G)}{bT}\!\sum_{r=1}^{T}(\vect{D}^{\!-1}\vect{A})^{r}\vect{D}^{\!-1}$ and $ \vect{L^{\prime}R^{\prime}}=\frac{\mathrm{vol}(G)}{bT}\! \vect{D}^{-1+\alpha}\vect{U}_k\vect{\Lambda}_k\left(\sum_{r=1}^{T}\vect{K}^{r-1}\right)\vect{U}_k^{\top}\vect{D}^{-1+\alpha}$, we have 
	\beq{
		\label{proof_eq5}
		\besp{
			&\Vert\vect{M}-\vect{L^{\prime}R^{\prime}}\Vert_F\\
			&=\frac{\mathrm{vol}(G)}{bT}\Vert\vect{D}^{-1/2}(\vect{X}-\vect{Y}+\vect{X}^2-\vect{Y}^2+\cdots+\vect{X}^{T}-\vect{Y}^{T})\vect{D}^{-1/2}\Vert_F\\
			&\leq \frac{\mathrm{vol}(G)}{bT}d_{min}^{-1}\sqrt{\sum_{i=1}^{n}\sigma_i(\vect{X}-\vect{Y}+\vect{X}^2-\vect{Y}^2+\cdots+\vect{X}^{T}-\vect{Y}^{T})}\\
			&\leq\frac{\mathrm{vol}(G)}{bT}\Vert\vect{X}-\vect{Y}+\vect{X}^2-\vect{Y}^2+\cdots+\vect{X}^{T}-\vect{Y}^{T}\Vert_F\\
			&\leq\frac{\mathrm{vol}(G)}{bT}(\Vert\vect{X}-\vect{Y}\Vert_F+\Vert\vect{X}^2-\vect{Y}^2\Vert_F+\cdots+\Vert\vect{X}^T-\vect{Y}^T\Vert_F).
    }}With Lemma~\ref{lemma4}, we have $\Vert\vect{X}-\vect{Y}\Vert_F\leq(1+\varepsilon)d_{min}^{-1+2\alpha}\sqrt{\sum_{j=k+1}^{n}|\lambda_j|^2}$. Therefore, 
	\beq{
		\label{proof_eq6}
		\besp{
			&\Vert\vect{X}^p-\vect{Y}^p\Vert_F\\
			&=\Vert\vect{X}^p-\vect{X}^{p-1}\vect{Y}+\vect{X}^{p-1}\vect{Y}-\vect{Y}^p\Vert_F\\
			&=\Vert\vect{X}^{p-1}(\vect{X}-\vect{Y})+(\vect{X}^{p-1}-\vect{Y}^{p-1})\vect{Y}\Vert_F\\
			&\leq\Vert\vect{X}^{p-1}\Vert_F\Vert\vect{X}-\vect{Y}\Vert_F+\Vert\vect{X}^{p-1}-\vect{Y}^{p-1}\Vert_F\Vert\vect{Y}\Vert_F.
	}}Considering eigen-decomposition of $\vect{G}\vect{H}\vect{G}^{T}=\vect{D}^{-1/2}\vect{A}\vect{D}^{-1/2}$, we  have eigenvalues $h_i\in[-1,1]$ according to~\cite{netmf}. Then,  $\Vert\vect{X}^{p-1}\Vert_F=\Vert\vect{G}\vect{H}^{p-1}\vect{G}^{\top}\Vert_F\leq n$ and $\Vert\vect{X}^p-\vect{Y}^p\Vert_F\leq n\Vert\vect{X}-\vect{Y}\Vert_F+\Vert\vect{X}^{p-1}-\vect{Y}^{p-1}\Vert_F\Vert\vect{Y}\Vert_F$.
	With Lemma~\ref{lemma3}, we derive
	\beq{
		\label{proof_eq7}
		\besp{
			\Vert\vect{Y}\Vert_F&=\Vert\vect{D}^{-1/2+\alpha}\vect{U}_k\vect{\Lambda}_k\vect{U}_k^{\top}\vect{D}^{-1/2+\alpha}\Vert_F
			\leq d_{min}^{-1+2\alpha}\Vert\vect{U}_k\vect{\Lambda}_k\vect{U}_k^{\top}\Vert_F.
	}}Considering $\vect{U\Lambda}\vect{U}^{\top}=\vect{D}^{-\alpha}\vect{A}\vect{D}^{-\alpha}$ with Lemma~\ref{lemma3}, we see that the $i$-th singular value of $\vect{U\Lambda}\vect{U}^{\top}$ satisfies
	\beq{
		\label{proof_eq8}
		\besp{
			\sigma_{i}(\vect{U\Lambda}\vect{U}^{\top})\leq d_{max}^{1-2\alpha}\sigma_{i}(\vect{D}^{-1/2}\vect{A}\vect{D}^{-1/2})\leq d_{max}^{1-2\alpha}.
    }}Combining (\ref{proof_eq7}) and (\ref{proof_eq8}), we have $\Vert\vect{Y}\Vert_F\leq \sqrt{k}(\frac{d_{max}}{d_{min}})^{1-2\alpha}\leq n(\frac{d_{max}}{d_{min}})^{1-2\alpha}$. Then $\Vert\vect{X}^p-\vect{Y}^p\Vert_F\leq n\Vert\vect{X}-\vect{Y}\Vert_F+\Vert\vect{X}^{p-1}-\vect{Y}^{p-1}\Vert_F\Vert\vect{Y}\Vert_F\leq n\Vert\vect{X}-\vect{Y}\Vert_F+n(\frac{d_{max}}{d_{min}})^{1-2\alpha}\Vert\vect{X}^{p-1}-\vect{Y}^{p-1}\Vert_F$.

	Combining (\ref{proof_eq5}) and (\ref{proof_eq6}), we derive
	\beq{
		\nonumber
        \besp{
			&\Vert\vect{M}-\vect{L^{\prime}R^{\prime}}\Vert_F\leq\frac{\mathrm{vol}(G)}{bT}\left(\Vert\vect{X}-\vect{Y}\Vert_F+\cdots+\Vert\vect{X}^T-\vect{Y}^T\Vert_F\right)\\
			&\leq\frac{(1+\varepsilon)d_{min}^{-1+2\alpha}\sqrt{\sum_{j=k+1}^{n}|\lambda_j|^2}((c^T-1)(1+\frac{n}{c-1})-nT)\mathrm{vol}(G)}{(c-1)bT},
	}}where $c=n(\frac{d_{max}}{d_{min}})^{1-2\alpha}$.
It is easy to observe that $\mathrm{trunc\_log}^{\circ}$ is 1-Lipchitz w.r.t. Frobenius norm~\cite{netsmf}. So, finally we have
    \beq{
		\label{proof_eq10}
		\besp{
			&\Vert f^{\circ}(\vect{M})-f^{\circ}(\vect{L^{\prime}R^{\prime}})\Vert_F=\Vert\mathrm{trunc\_log}^{\circ}(\vect{M})-\mathrm{trunc\_log}^{\circ}(\vect{L^{\prime}R^{\prime}})\Vert_F\\
			&\leq\frac{(1+\varepsilon)d_{min}^{-1+2\alpha}\sqrt{\sum_{j=k+1}^{n}|\lambda_j|^2}((c^T-1)(1+\frac{n}{c-1})-nT)\mathrm{vol}(G)}{(c-1)bT}\\
			&=\frac{(1+\varepsilon)d_{min}^{-1+2\alpha}B}{(c-1)bT},
    }}where $B=\sqrt{\sum_{j=k+1}^{n}|\lambda_j|^2}((c^T\!-\!1)(1+\frac{n}{c\!-\!1})\!-\!nT)\mathrm{vol}(G)$ and $c=n(\frac{d_{max}}{d_{min}})^{1-2\alpha}$.
\end{proof}



\ifCLASSOPTIONcaptionsoff
  \newpage
\fi



\bibliographystyle{IEEEtran}
\bibliography{IEEEabrv}
%

%

\hide{
\begin{IEEEbiography}{Michael Shell}
Biography text here.
\end{IEEEbiography}

\begin{IEEEbiographynophoto}{John Doe}
Biography text here.
\end{IEEEbiographynophoto}


\begin{IEEEbiographynophoto}{Jane Doe}
Biography text here.
\end{IEEEbiographynophoto}
}



